\documentclass{elsart}
\usepackage{epsfig}
\usepackage{amssymb}

\begin{document}

\newcommand{\be}{\begin{equation}}
\newcommand{\ee}{\end{equation}}
\newcommand{\bea}{\begin{eqnarray}}
\newcommand{\eea}{\end{eqnarray}}
\newcommand{\rr}{{\bf r}}
\newcommand{\rt}{\rr^\perp}
\newcommand{\kk}{{\bf k}}
\newcommand{\kkt}{\kk^\perp}
\newcommand{\p}{{\bf p}}
\newcommand{\q}{{\bf q}}
\newcommand{\qt}{\q^\perp}
\newcommand{\X}{({\bf x})}
\newcommand{\Y}{({\bf y})}
\newcommand{\x}{{\bf x}}
\newcommand{\ff}{{\bf f}}
\newcommand{\uu}{{\bf u}}
\newcommand{\xx}{{\bf x}}
\newcommand{\EE}{{\bf E}}
\newcommand{\VV}{{\bf V}}
\newcommand{\y}{{\bf y}}
\newcommand{\U}{{\bf u}}
\newcommand{\w}{{\bf \omega}}
\newcommand{\D}{{\bf \nabla}}
\newcommand{\W}{\omega_\kk}
\newcommand{\za}{\alpha}
\newcommand{\zb}{\beta}
\newcommand{\zd}{\delta}
\newcommand{\zg}{\gamma}
\newcommand{\zl}{\lambda}
\newcommand{\zs}{\sigma}
\newcommand{\zt}{\tau}
\newcommand{\zN}{I\hskip-3.4pt N}
\newcommand{\zR}{I\hskip-3.4pt R}
\newcommand{\zw}{\omega}
\newcommand{\zC}{{\mathbb C}}
\newcommand{\OM}{({\bf ***\ldots***})}
\newcommand{\EM}{({\bf $\leftarrow$***})}
\newcommand{\BM}{({\bf ***$\rightarrow$})}

\begin{frontmatter}

\title{Lyapunov spectra and nonequilibrium ensembles equivalence  \\ 
in 2D fluid mechanics}

\author[GG]{Giovanni Gallavotti}
\ead{gallavotti@roma1.infn.it}
\author{Lamberto Rondoni\corauthref{LR}} 
\ead{rondoni@polito.it}
\author[ES]{Enrico Segre} 
\ead{segre@athena.polito.it} 

\address[GG]{Fisica, Universit\`a di Roma ``La Sapienza'', P.le Moro 2,   
             I-00185 Roma, Italy}
\address[LR]{DIMAT, Politecnico di Torino, Corso Duca degli Abruzzi 24,
                 I-10129 Torino, Italy and INFM}

\address[ES]{DIASP, Politecnico di Torino, Corso Duca degli Abruzzi 24,
                 I-10129 Torino, Italy}

\begin{abstract}
We perform numerical experiments to study the Lyapunov spectra of
dynamical systems associated with the Navier--Stokes (NS) equation in
two spatial dimensions truncated over the Fourier basis. Recently new
equations, called GNS equations, have been introduced and conjectured
to be equivalent to the NS equations at large Reynolds numbers.  The
Lyapunov spectra of the NS and of the corresponding GNS systems
overlap, adding evidence in favor of the conjectured equivalence
already studied and partially extended in previous papers.  We make
use of the Lyapunov spectra to study a fluctuation relation which had been
proposed to extend the ``fluctuation theorem'' to strongly dissipative
systems. Preliminary results towards the formulation of a local
version of the fluctuation formula are also presented.
\end{abstract}

\end{frontmatter}

%
\section{Introduction}
The idea of viewing turbulent fluid motions as trajectories in the
phase space of a chaotic dynamical system is a long standing one.  In
particular, treating turbulence in terms of invariant probability
distributions on the phase space of the dynamical system given by the
Navier-Stokes (NS) equation for a fluid, and asserting that the
distributions ought to have special properties, is an idea which dates back
to the early 1960's: it has been formalized a little later,
see\cite{Ru78}.\footnote{Quoting \cite{BJPV98}, ``It is widely
accepted that the statistical properties of a turbulent flow can be
described by a dynamical evolution on a strange attractor in a
high-dimensional phase space. An important characterization of the
ergodic properties of the attractor is given by the Lyapunov
spectrum.''}

Mathematical and numerical studies have been performed, at first for
the purpose of relating fluid dynamical parameters, such as the Reynolds
number or the amplitude of external forcing, to the fractal dimension
of the attractor\footnote{It is important, when discussing fractality,
to distinguish between {\it attractor} and {\it attracting set}.  An
attracting set is a {\it closed set} such that all points in its
vicinity get as close as one wants to the set as time tends to
infinity and, {\it furthermore}, such that none of its closed subsets have
the same property. The notion of attractor, that none of its closed
subsets have the same property. The notion of attractor is associated
with a method of choice of initial data in the vicinity of an
attracting set. Suppose that the data are randomly selected with a
distribution $\mu_0$ and, with $\mu_0$--probability $1$, their
asymptotic motion is described by a probability distribution $\mu$ on
the attracting set, in the sense that time averages can be obtained
simply by integration with respect to $\mu$. Then, {\it any} subset of
the attracting set with $\mu$--probability $1$ {\it and} smallest
Hausdorff dimension is called ``attractor'' for the random choices
with distribution $\mu_0$, and $\mu$ is called the {\it statistics of
$\mu_0$}. Quite commonly, the attracting set is not fractal, and it
might even coincide with the whole phase space, while the attractor
for (some) $\mu_0$ is fractal and with dimension lower than that of
the attracting set. We recall that the special statistics generated on
an attracting set from data randomly chosen with a probability
distribution $\mu_0$ absolutely continuous with respect to the volume
is called a ``{\it SRB distribution}'', which is usually unique: in
this paper we shall, as usual, only consider the SRB statistics.}. In
this way the onset of turbulence has been quite well understood, along
with the discovery that the relevant physical phenomena are far more
interesting and complex than previously anticipated,
see\cite{GraLeo91,Lee95,EcRu,Lie84,Ru84}.

However, understanding developed turbulence remains a most challenging
problem and it is not clear how to use the basic ideas of Lorenz,
Ruelle and Takens in this context.  The approach to the analysis is
very often based on numerical experiments in which simulations of
reduced models of turbulence are considered. For instance the shell
models of Yamada and Okhitani \cite{YamOhk88,YamOhk98,BPPV} have been
studied in great detail with the aim of finding or checking
connections between Lyapunov spectra and the turbulent cascade.

Another viewpoint \cite{GAFLU,Galibro-i} stems from the well known fact that
macroscopic irreversibility of particle systems can be obtained from
microscopic reversible dynamics, provided the ratios between the microscopic
and the macroscopic time and length scales are sufficiently large. In that
case, in fact, the microscopic reversible equations of motion are
``equivalent'' to dissipative equations of motion at the macroscopic
level.

Assuming that forced turbulence is described by the (irreversible) NS
equation, a different (reversible) equation called the ``Gaussian
Navier Stokes'', or GNS, equation was proposed to describe
the same physical phenomena \cite{GAFLU,Galibro-i}. In the GNS equation
the usual viscosity constant is replaced by a fluctuating term, which
makes the equation itself {\em time reversal invariant} and keeps the
total energy exactly constant.  Equivalence should occur when the
constant energy of the GNS evolution equals the average energy
in the NS evolution provided the Reynolds number is large: in a
way which is similar to the equivalence between microcanonical and
canonical ensembles in statistical mechanics\cite{GG-SMbook}, with the
large volume limit replaced by the limit of large Reynolds number
$R$. The conjecture states that the forward time statistics of the NS
and of the GNS models of the same fluid are the same, for what
concerns ``local observables'', {\it i.e.}  for the time averages of
functions of the velocity field which depend only on a small number
of its Fourier components physically corresponding to observables on
the inertial scale.

The idea, if correct, could be used to analyze phenomena which would
not be easily understood in terms of the NS equation. For instance,
time reversal invariance and chaoticity of the GNS dynamics imply that
the fluctuations of certain quantities verify a relation (cf.\
Ref.\cite{GalCoh95}), here referred to as ``fluctuation relation''
(FR), which might eventually be amenable to experimental verification
as attempted in Refs.\ \cite{CL,GGK}. It should be remarked that the
possibility of using several different representations of the
statistical properties of a turbulent fluid flow, which result
equivalent from a practical point of view, was considered already in
previous works, see, e.g.\ Refs.\cite{Germano,MenKa,SheJack}, and is
still under investigation \cite{LK01}.

Of course despite the expected equivalence, the NS and GNS
representations of a turbulent fluid remain different, similarly to
the microcanonical and the canonical ensembles descriptions of
equilibrium statistical mechanics: although ``equivalent in the
thermodynamic limit'' they differ in many respects ({\it e.g.} in the
distribution of fluctuations of the total energy).

\section{NS and GNS equations. Aims of our analysis. \label{secEqNS} }
Suppose that the side of the periodic cell is $L$, and that the
forcing term $\ff$ is the product of a dimensional parameter $F$ times
a fixed function. We shall suppose the Fourier components $F_\kk$ to
have modulus $1$ and, usually, to vanish for all but one value of
$\kk$.\footnote{The cases of Fig.\ref{sovra18} are the only exceptions
considered here.}
Then, in the NS case with viscosity $\nu$, we define
the {\it Reynolds number} $R$ to be

\be 
R^2=FL^3\nu^{-2}, \qquad {\rm or} \qquad R=\nu^{-1}{\sqrt{F L^3}}
\label{Rey} ~.
\ee
and we shall consider the NS equations in dimensionless form.  Here,
the lengths are measured in units of $L/2\pi$ and time in units of
$L^2/\nu$, so that the change of variables $\uu(x,t)=V {\bf
u}^g(\xx/L,Ct)$ with $V=FL^2/\nu, C=\nu/L^2$, yields (redefining
simply by $\uu$ the dimensionless field $\uu^g$ for notational
convenience) the {\it dimensionless NS equations}
\be
\dot \uu+R^2 (\uu\cdot\partial) \uu=\Delta \uu + \ff - \partial p ~, \quad
\partial \cdot \uu = 0
\label{5}
\ee
with $\uu$ periodic in the position coordinates with period $2\pi$ and
{\it divergenceless} ({\it i.e.} $\uu$ represents an incompressible
flow).  Correspondingly, the GNS equations will be defined by 
\be \dot
\uu+R^2 (\uu\cdot\partial) \uu=\alpha\, \Delta \uu + \ff - \partial p
~, \quad \partial \cdot \uu = 0
\label{6}
\ee
where $R$ is a free parameter and $\za$ is defined so that the
solutions of the equation (\ref{6}) conserve $Q_0=\int \uu^2 d\xx$,
{\it i.e.} twice the total kinetic energy.

In order to compare motions that the GNS equations generate with
(suitably selected) motions of the NS fluid, the constant $Q_0$ has to
be chosen.  For a given $R$ and taking as reference the NS equation,
we call $\langle Q_0\rangle$ the infinite time average of $Q_0$
produced by the NS evolution. We shall compare, according to the
analysis of \cite{Gal97,RonSeg99}, the solutions of the NS equation
with those of the GNS equations with the conserved quantity $Q_0$ set
equal to $\langle Q_0\rangle$.

We write $\uu(\xx)=\sum_\kk e^{i\kk\cdot\xx} u_\kk \kk^\perp/|\kk|$,
where the scalar term $u_\kk$ obeys $u_\kk = -\overline{u_{-\kk}}$,
and $\kk^\perp = (k_2,-k_1)$ if $\kk = (k_1,k_2)$.  Hence
$Q_0=(2\pi)^2\sum_\kk |u_\kk|^2$ (and the total vorticity is
$Q_1=(2\pi)^2\sum_\kk |\kk|^2|u_\kk|^2$).  If we define the {\it
effort} ${\bf G}_1$ as

\bea
{\bf G}_1 &=& \big(\Delta^{-1}
({\bf a}-\ff-\partial p),({\bf a}-\ff-\partial p)\Big) = \nonumber\\
&=&(2\pi)^2\sum_\kk 
\frac{\big( \dot u_\kk+iR^2\sum_{{\bf j}+{\bf h}=\kk}
\frac{({\bf j}^\perp \cdot {\bf h})({\bf h}\cdot\kk)}
{|{\bf j}||{\bf h}||{\bf k}|} u_{\bf j} u_{\bf h}- f_\kk\big)^2}{\kk^2}
\label{sforzo}
\eea
where $(\cdot,\cdot)$ denotes the usual scalar product in $L_2(d\xx)$,
and ${\bf a}=d\uu/dt=\partial_t\uu+R^2(\uu\cdot\partial)\,\uu$ is the
acceleration field, Eq.(\ref{6}) can be obtained from Gauss' least
effort principle by considering an Euler fluid subject to the
constraint of constant $Q_0$, {\it i.e.} constant total energy.  In
this case the acceleration ${\bf a}$ of the velocity field $\uu$
minimizes the effort ${\bf G}_1$ among all possible accelerations
compatible with the incompressibility and with the constraint.  If we
write Eq.(\ref{6}) in terms of Fourier components we find \be
\dot{u}_\kk= -i R^2\sum_{{\bf j}+{\bf h}=\kk} \frac{({\bf j}^\perp
\cdot {\bf h})({\bf h}\cdot\kk)} {|{\bf j}||{\bf h}||{\bf k}|} u_{\bf
j} u_{\bf h} - \alpha \kk^2 u_\kk + f_\kk
\label{NSk0}
\ee 
where $\alpha$ is given by
\be
\alpha=\frac{\sum_\kk f_\kk \overline u_\kk}{\sum_\kk
|\kk|^2|u_\kk|^2}\label{alfa} ~.
\ee
Of course the fact that the equations above can be obtained from a
variational principle might be merely accidental, and in itself does not
justify their use. However in
Refs.\cite{GAFLU,Gal97,RonSeg99,Galibro-i} it has been argued that they
should give a description of the fluid motions which in the
large Reynolds number regime is as accurate as the NS equations
themselves.  More generally, one can consider an Euler fluid subject
to a Gaussian constraint on $Q_m$ with effort function ${\bf
G}_\ell$, $\ell\le m+1$, given by

\bea
Q_m &=& (2\pi)^2\sum_\kk |\kk|^{2m} |u_\kk|^2, \nonumber \\
{\bf G}_{\ell,m} &=&
\big((\Delta)^{-(\ell-m)}({\bf a}-\ff-\partial p),({\bf a}-\ff-\partial
p)\big) \label{alfax}~.
\eea
This leads to equations like (\ref{6}) with the Laplace operator
replaced by $\Delta^\ell$, and with $\alpha$ replaced by a new
coefficient $\beta_{\ell,m}$ that can be easily computed by imposing
that $Q_m$ remains exactly constant. We call the
resulting equations ``{\it G-hyperviscous}'' equations while we call
``{\it hyperviscous}'' the corresponding equations in which
$\beta_{\ell,m}$ is replaced by a constant. The above case (\ref{NSk0})
corresponds to $\ell=1,m=0$. The special cases $\ell=0$, $m=0$ and
$\ell=1,m=1$ have been considered in \cite{GAFLU} and the general
hierarchy of the hyperviscous equations has been considered and
partially studied in \cite{RonSeg99}. The case $\ell=1,m=1$, for instance, 
yields:
\be \beta_{1,1}=\frac{\sum_\kk |\kk|^2 f_\kk \overline u_\kk} 
{\sum_\kk |\kk|^4|u_\kk|^2} ~.
\label{alfa1} 
\ee
A general feature of the G-hyperviscous equations is that they
generate {\it reversible dynamics} in the sense that the velocity
reversal operation $I u_\kk=-u_\kk$ anticommutes with the time
evolution map $S_t$ (the mapping of an initial datum $\uu$ into its
value $S_t\uu$ at time $t$) in the sense that

\be
\label{tr} 
I\cdot S_t=S_{-t}\cdot I ~.
\ee
\vskip3truemm
Note that here we call GNS equations the case $\ell=1,m=0$ for
consistency with \cite{RonSeg99} because we shall often
refer to Ref.\cite{RonSeg99}, although in \cite{GAFLU} this name
was used for the case $\ell=1,m=1$.
\vskip3truemm

In this paper we study via computer simulations:

\begin{itemize}
\item{(1)} Relations between the Lyapunov spectra of corresponding NS
and GNS models. Equivalence between the Lyapunov spectra is not part
of the earlier proposals on equivalence because the Lyapunov exponents
are not ``local observables'': however it seems possible that they
reflect properties of the individual modes of the velocity
field \cite{hopo} so that it is interesting to test whether equivalence
extends to them.  In the models that we consider the Lyapunov spectra
appear to be essentially {\it identical} within the errors.

\item{(2)} Fluctuation properties of the observable defined by the
phase space contraction in the reversible GNS equation and its
relation with the Lyapunov spectra following the proposal in
\cite{GAFLU}. The relation proposed in \cite{GAFLU} is, however, based
on several assumptions which we cannot check directly.  Therefore we
have measured quantities for which predictions are available as
consequences of the mentioned assumptions, finding that the
predictions match with the observations. This lends support to the
idea that the property called ``{\it axiom C}\,'' as well as the
assumptions on the structure of the attracting sets discussed in
\cite{GAFLU,BG} may be actually verified in the models considered
here.

\item{(3)} The equivalence analysis can be extended to involve not
only NS and GNS equations but also to several members of the hierarchy
of hyperviscous and G-hyperviscous equations. We have made a few tests
analogous to those made in the case of the NS and GNS equations, and
some of the results are briefly presented below.

\item{(4)} Local fluctuations. The theory of the fluctuations of
$\alpha$ considered in item (2) above concerns the global phase space
contraction rate, which in our models is extensive. Extensivity makes
it very difficult to measure the fluctuations of $\za$ in systems with
a large number of modes ({\it i.e.} many degrees of freedom). To
overcome this difficulty, we define and study the ``local phase space
contraction'' related to a portion of the volume occupied by the
fluid, where fluctuations are strongly enhanced.

The investigation is important because in comparing fluctuations of
phase space contraction between corresponding NS and GNS evolutions
one must look at smaller subsystems. For instance, in the (truncated)
NS equation there is no fluctuation of the {\em total} phase space
contraction rate, because the friction coefficient is constant: the
phase space contraction rate ({\it i.e.} the divergence of the
equations of motion) is simply $-2\sum_{|\kk|<N} \kk^2$ if $N$ is the
cut off parameter.

Furthermore, in real experiments, one is forced to look at the
fluctuations within small portions of fluid in order to obtain good
statistics \cite{CL,GGK}. Our results on this matter are only 
preliminary, but encouraging so that we report them in the Appendix.
\end{itemize}

We integrated Eq.(\ref{NSk0}) by means of a pseudospectral code
truncating it to a small number $K$ of Fourier modes, $u_\kk$, of the
velocity field. The computation of the entire Lyapunov spectrum is,
generally, expensive (in terms of CPU time), and testing the validity
of the FR (for the GNS equation) requires high statistics on the
fluctuations of the phase space contraction rate $\alpha$.  Thus we
have been forced to adopt a ``low resolution'' approximation of the
dynamics in Fourier space: i.e.\ to take
$\kk,\p,\q\in[-(N-1),...,(N-1)]^2$, where $N$ is a small truncation
value. In this way, the number of independent real components of the
velocity field is $2K$ with $2K=4N(N-1)$ not counting the components
corresponding to $\kk=\bf0$ which, being constants of motion, have
been fixed to be zero. We say that the equations are ``truncated at
$K$ modes'' calling ``mode $\kk$'' a (complex) Fourier component
$u_\kk$ of the velocity field.  In our cases $2K$ ranges between $24$
and $168$: this corresponds to $N-1$ between $2$ and $6$. Given our
Reynolds numbers, $K$ is much less than the value it should have
(which we estimate to be about one order of magnitude higher) in order
to interpret the results as relevant for developed turbulence (albeit
in two dimensions), consistently with the theories of homogeneous
turbulence. However, our present computer facilities make it
impossible for us to reach much higher numbers of modes with the
precision demanded by our analysis.

\vskip 5pt
\vskip 3mm

\section{Extension of equivalence between the NS and GNS dynamics to 
Lyapunov spectra}
The equivalence principle introduced in Refs.\cite{GAFLU,Gal97} yields
quantitative relations between certain observables, which can be
tested in numerical experiments. The principle states that NS and GNS
equations should have same statistics for the ``local
observables''\footnote{{\it i.e.}  for the time averages of functions
of the velocity field which depend only on a finite number of its
Fourier components or, more generally and more physically, which
depend only on Fourier components of modes with ${\bf k}$ in the
inertial range.}, when the limit of large Reynolds number is taken and
the average phase space contraction of the GNS evolution equals the
(constant) phase space contraction rate of the NS evolution.

The phase space contraction rate associated with a differential
equation is the divergence of the equations of motion: it can be
straightforwardly computed from Eqs.\ (\ref{NSk0}),(\ref{alfa}) to be
$\sigma= (-1+2\sum_{|k_i|<N}1)\,\cdot\,\alpha$ in the case $\ell=0$
and $m=0$ (which we call GED case), and \be \sigma^{GNS}=2
\left(\sum_i |k_i|^2 - {Q_{2}\over Q_{1}} \right)\,\cdot\,\alpha+
{\sum_{|k_i|<N} \kk^{2}f_{k_i}\overline u_{k_i}\over Q_{1}}
\label{pscr}
\ee
in the GNS
case, i.e.\ for $\ell=1,m=0$ (see 
Eq.(\ref{alfax}))\footnote{In experiments on
real fluids this quantity must be associated with some physical
observable (a very delicate matter, \cite{CL,GGK})}. Denoting by
\be
\zs^{NS}=2\sum_{|k_i|<N} |k_i|^2=M
\label{MNS}
\ee
the (constant) phase space contraction rate of the NS evolution, and by 
\bea
\langle \sigma^{GNS} \rangle &=&
\Big\langle
 2 \left(\sum_i |k_i|^2 - {Q_{2}\over Q_{1}} \right)\,\cdot\,\alpha+
  {\sum_{|k_i|<N} \kk^{2}f_i\overline u_{k_i}\over Q_{1}} 
 \Big\rangle= \nonumber \\ 
&=&M\cdot\langle \alpha \rangle +o(M)
\label{MGNS}
\eea
the average phase space contraction rate of the GNS evolution, the 
equivalence conjecture studied in \cite{RonSeg99} can be stated as follows.

\vskip 3mm

\noindent
{\bf Equivalence Conjecture (EC).} {\it The stationary probability
distributions on the phase space associated with the NS equations and
the GNS equations are {\em equivalent} in the limit of large Reynolds
number, provided the value of $Q_0$ is chosen so that $\zs^{NS}$ and $
\langle\zs^{GNS}\rangle $ coincide or, equivalently, provided the
value of $Q_0$ is chosen so that $\langle\alpha \rangle $ equals
$1$.\footnote{{\it i.e.} equals the viscosity in our dimensionless
units.}  }
\vskip 3mm

\noindent
The idea behind this conjecture is that if the fluctuations of
$\alpha$ occur on time scales which are short compared with the
macroscopic observation time scales, and if $\langle \alpha \rangle=1$,
the macroscopic observables should be equally well computed from
the NS and the GNS dynamics. The large $R$ limit is, then, required to
ensure that the fluctuations of $\alpha$ do really occur on
sufficiently short time scales. The results of \cite{RonSeg99} suggests that
this is the case. 

To obtain such results, the steady state of the NS evolution was first determined
by following during a long time $T$ the approach
of the time average of $Q_0$ to its asymptotic value $\langle
Q_0\rangle$. At time $T$, the evolution was continued with the GNS
dynamics with initial condition taken to be the vorticity field
produced by the NS dynamics at time $T$.  It was then observed that
$\langle \zs^{GNS} \rangle / \zs^{NS} \approx \langle\alpha\rangle$,
and that $\langle\alpha\rangle$ differ little from $1$ already at
small $N$ (of order $O(10^1)$ and even less) {\em as long as the
initial (fixed) value $Q_0^{GNS}$ of $Q_0$ was close to its NS
time average $\langle Q_0\rangle$}.  Furthermore the details of the
initial condition for the GNS evolution did not affect {\em too much}
the statistical results produced by the GNS dynamics.  In other words,
it appeared that only the parameter $Q_0^{GNS}$ needs to be adjusted
to obtain equivalent NS and GNS evolutions. Although the meaning of
``{\em too much}'' and {\em `` close''} was not sharply quantified in
\cite{RonSeg99}, the results suggested the validity of an EC also
in the following alternative sense:

\vskip3mm

\noindent
{\bf Equivalence Conjecture (EC$'$).} {\it The stationary phase space
probability distributions of the NS equations and of the GNS equations
are {\em equivalent} in the limit of large Reynolds number,
provided the energy $Q_0$ is fixed in the GNS evolution to
coincide with the average value $\langle Q_0 \rangle$ of the NS
evolution at the same (large) $R$.  }

\vskip 3mm \noindent In this paper, we first investigate the
possibility of extending the EC,EC$'$ to the equivalence of the
Lyapunov spectra.  We evaluate the Lyapunov exponents of the NS
equation by performing a long numerical simulation from time $0$ up to
time $T$, yielding the NS Lyapunov spectrum and the average value
$\langle Q_0 \rangle$, with an error bar $\zd Q_0$.  From
time $T$, we continue the NS evolution until $Q_0$, which fluctuates, equals
$\langle Q_0 \rangle$. This usually takes a short time $t_s$. Then, at
time $T+t_s$, we switch to the GNS evolution and compute the
corresponding GNS Lyapunov spectrum. The reason to adopt this
procedure, rather than just taking a random initial field with
$Q_0^{GNS}=\langle Q_0 \rangle$ for the GNS evolution, is that {\it a
priori} the attractor might be not unique, and different initial
fields may be evolved by the GNS dynamics towards attractors different
from the one found by following the NS evolution. In other words, the
NS and the GNS statistical properties may not always result comparable
because of possible hysteresis phenomena. Our procedure was originally
designed to avoid this problem, although it turned out subsequently
that our choice of parameters yields no hysteresis, as discussed in
Ref.\cite{GVR}.

\begin{figure}
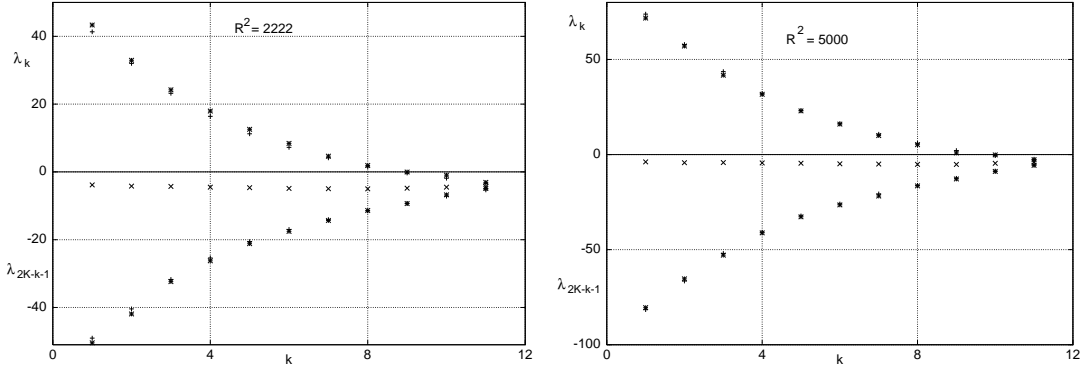
\centerline{\hbox{
\epsfig{figure=n0015.pst,width=5.0cm,angle=-90}
\epsfig{figure=n001.pst,width=5.0cm,angle=-90}
}} 
\caption{\small Lyapunov spectra for the NS runs with normal viscosity
($\ell=1$) at $R^2=2222$ (left) and $R^2=5000$ (right), and
corresponding GNS runs with constrained energy $Q_{0}$.
The $2K-2$ nontrivial exponents are drawn by associating each value of the
abscissa $k=1,2,\ldots,K-1$ with the $k$--th largest exponent
$\lambda_k$ and the $k$--th smallest exponent
$\zl_k'=\lambda_{2K-k-1}$. 
The symbols ``$+$'' refer to the NS spectra, the ``$*$'' to the GNS
spectra, and the ``$\times$'' to the sums $(\lambda_k+\lambda'_k)/2$
for the NS case.
There is no ``pairing'' of the exponents
to a common average value, unlike the cases of isokinetic Gaussian
systems \cite{DetMor96}.  
Essentially identical results have been obtained by studying the ED and GED
equations.} 
\label{n002n001} 
\end{figure}

The Lyapunov spectra produced by the NS and GNS dynamics in this way 
appear {\em equivalent}, meaning that the Lyapunov spectra of
the NS and GNS dynamics appear to become identical for large $R$ and
$K$. At finite $R$ and $K$, instead, the sums of the Lyapunov
exponents should be related by (cf.\ eqs.(\ref{MNS},\ref{MGNS})) 
\be \sum_{i=1}^K \zl_i^{GNS} = \sum_{i=1}^K \zl_i^{NS} + o(M) ~, \quad
\mbox{where } ~~~ \sum_{i=1}^K \zl_i^{NS} = - M.
\label{GNStoNS}
\ee 
The typical situation is represented in Fig.\ \ref{n002n001} where the
$2K-2$ nontrivial exponets are drawn. Note that two exponents are {\it
a priori} known to vanish in the GNS case: one corresponding to
the direction of motion and one to the direction of the conserved
quantity; in the NS case one is {\it a priori} known to vanish,
corresponding to the direction of motion. However in all cases the NS
evolution shows another exponent very close to zero in agreement with
the equivalence conjecture and we have taken it to be zero in
drawing the figures 1 and 2 below, see also footnote 11 below.

The Lyapunov exponents are obtained by the method of Ref.\cite{BGGS}
over times $T$ of the order of up to $10^5$ Lyapunov time
units.\footnote{We define the Lyapunov time unit as $1/\zl_{max}$,
where $\zl_{max}$ is the largest Lyapunov exponent.}  Error bars do
not appear, since they are not larger than the size of the symbols in
the figures, as explained below in Section 5.

Consistently with the EC$'$, and up to numerical errors, we observed
that the viscosities and the average enstrophies of the NS and GNS
evolutions with $2K=24$ result equal if the average energies are
equal. The numerical errors we refer to can be attributed to our
approximate knowledge of the average of $Q_0$ (which is affected by an
error bar $\zd Q_0$); and to the finite values of $R$,$N$, and $T$
(which, in turn, can only be reached through a sequence of time steps
of finite size $h$). The corresponding deviations from the conjectured
equalities of the viscosities and of the average enstrophies are
measured by the terms $\triangle\alpha = (\langle \za \rangle - 1)$
and $\triangle Q_1 = |\langle Q_1 \rangle_{NS} - \langle Q_1
\rangle_{GNS}|/\langle Q_1 \rangle_{NS}$ in Tables \ref{newtab} and
\ref{newGED}, where the subscripts $NS$ and $GNS$ indicate the two
relevant kinds of dynamics.
\begin{table}
\begin{center}
\begin{tabular}{
||r     |c     |      c  |            c |           c||} \hline \hline
$ R^2 $ &$\zd Q_0/\langle Q_0 \rangle_{NS}$
&$\triangle \alpha$&$\triangle Q_1$& $o(M)/M$ \\
\hline				   
800 & 0.005 & 0.030 & 0.053 & 0.068 \\
\hline				   
1250 & 0.020 & 0.018 & 0.062 & 0.057 \\
\hline				   
2222 & 0.002 & 0.039 & 0.058 & 0.077 \\
\hline				   
4444 & 0.050 & 0.021 & 0.093 & 0.059 \\
\hline				   
5000& 0.010 & 0.008 & 0.058 & 0.033 \\
\hline \hline
\end{tabular}
\caption{\small Equivalence of NS and GNS dynamics, i.e.\ with
$\ell=1$ and $m=0$, for different Reynolds numbers. The last column
gives the relative difference of the computed sums of the NS and GNS
Lyapunov exponents (cf.\ eq.(\ref{GNStoNS})).}
\label{newtab}
\end{center}
\end{table}
\begin{table}
\begin{center}
\begin{tabular}{
||r     |c     |      c  |            c |           c||} \hline \hline
$ R^2 $ &$\zd Q_0/\langle Q_0 \rangle_{NS}$
&$\triangle \alpha$&$\triangle Q_1$& $o(M)/M$ \\
\hline				   
408  & 0.002 & 0.039 & 0.001 & 0.060 \\
\hline				   
800 & 0.002 & 0.038 & 0.001 & 0.067 \\
\hline				   
2222 & 0.003 & 0.003 & 0.002 & 0.042 \\
\hline \hline
\end{tabular}
\caption{\small Same as in Table \ref{newtab} but for the ED and GED case
with $\ell=0$ and $m=0$.}
\label{newGED}
\end{center}
\end{table}

Further computations of the Lyapunov spectra were performed for $2K=168$,
starting from random initial data, evolving with the NS
evolution, and then switching to a GNS evolution, as above. In this case,
however, the fixed quantity $Q_m$ of the GNS evolution was the energy $Q_0$ in
one instance, the enstrophy $Q_1$ in another instance, and the palinstrophy $Q_2$
in the last instance.
The motivation of this computation was to
check the validity of the conjecture, proposed and partially tested in
\cite{RonSeg99}, that ``several'' pairs of equations of the hierarchies
of hyperviscous and G-hyperviscous equations, with $\ell \le m+1$, are
equivalent to each other in the limit of large Reynolds numbers.

\begin{figure} \centering
\psfig{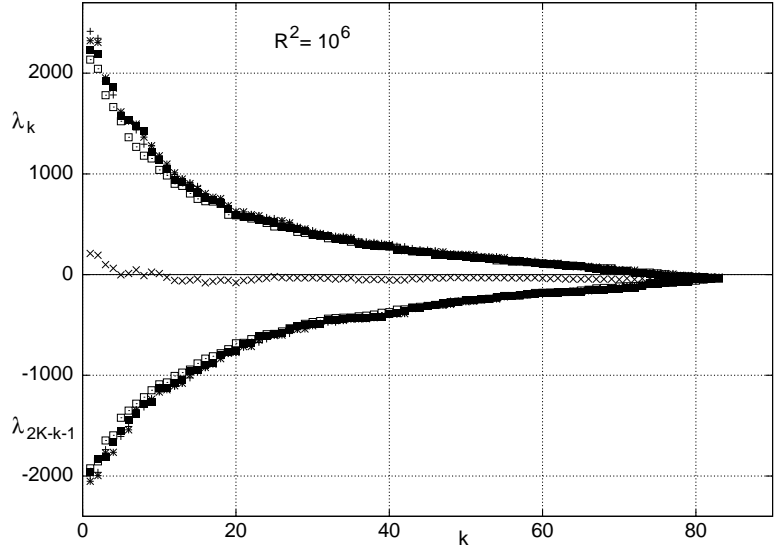}
\caption{\small Lyapunov exponents for one NS run with normal
viscosity ($+$) at $N=7$, $R^2=10^6$, and forcing on modes $(4,-3),
(3,-4)$. The $2K-2$ nontrivial exponents are drawn as in Fig.\
\ref{n002n001}. The middle line ($\times$) is the graph of
$(\lambda_k+\lambda_k')/2$.  The other symbols refer to the
corresponding GNS runs with fixed energy ($\ast$) {\it i.e.}
$m=0,\ell=1$, fixed ``enstrophy'' ($\boxdot$) {\it i.e.}  $m=1,\ell=1$,
or fixed ``palinstrophy'' ($\blacksquare$) {\it i.e.}  $m=2,\ell=1$.
The spectra are obtained from runs of different lengths with $T \in
[125,250]$, in units of $1/\zl_{max}$, $\zl_{max}$ being the largest
Lyapunov exponent. The overlap of the four spectra (albeit
approximate, because of the uncertianties $\zd Q_0, \zd Q_1, \zd Q_2$)
reflects the validity of the extension of the EC,EC$'$ to the whole
spectrum and to different members of the hierarchy of equations.  The
error bars, see Sec. 5, can be identified with the size of the
symbols.}
\label{sovra18}
\end{figure}

If the EC can be extended to such general cases and to cover the
Lyapunov spectra, the spectra should overlap (for large $R$).
Considering the fact that the error bars $\zd Q_0, \zd Q_1, \zd Q_2$
on $\langle Q_0 \rangle, \langle Q_1 \rangle, \langle Q_2 \rangle$
in these cases were larger than in the cases with $2K=24$ (the largest
being $\zd Q_1 \approx 0.09 \langle Q_1\rangle$), our
results for $\ell=1$ and $m=0,1,2$, seem to confirm this property.
This is illustrated by Fig.\ref{sovra18}
which shows the equivalence, within our numerical accuracy, of
the computed Lyapunov spectra of the NS and GNS systems.
A similar equivalence was observed in
several other cases which we do not describe here for brevity.  This
reinforces the previous results of \cite{RonSeg99}, and provides a
heuristic motivation for the validity of the conjectures EC, EC$'$.

In Fig.\ref{n002n001} and Fig.\ref{sovra18} two exponents for the GNS
equation vanish exactly reflecting that $Q_0$ is a constant of motion
and neither expands nor contracts in the direction of the flow.
There is a third exponent close to $0$ which, however, seems to
be numerically distinct from $0$. In the NS cases only one exponent is
certainly $0$ (corresponding to the flow direction) but there are two
other exponents which are numerically very close to $0$, see also
footnote${}^{11}$ below.

\section{Time reversal symmetry, axiom C and the fluctuation theorem}
A fluctuation theorem (FT), \cite{GalCoh95-0}, for the fluctuations of
the phase space contraction rate $\sigma^{GNS}(\uu)$ in our GNS
truncated systems can be derived under the assumption that the Chaotic
Hypothesis (CH) is verified by the GNS dynamics at the considered
Reynolds number, \cite{GalCoh95}. The CH can be formulated as follows:

\vskip 5pt
\noindent
{\bf Chaotic Hypothesis (CH).} {\it A chaotic many-particle system or
{\em fluid} in a stationary state can be regarded, for the purpose of
computing macroscopic properties, as a smooth dynamical system with an
attracting set on which the system is a smooth transitive Anosov
system.}
\vskip 3mm

This means that the attracting set can be considered, {\it for
practical purposes}, to be a connected ``smooth surface'', on which the
dynamics is hyperbolic and in fact verifies the Anosov property, see
\cite{Galibro-i} for a mathematical definition.  In
other words, one supposes that the attracting set is a smooth surface
possibly of dimension lower than that of phase space, thinking that
its fractality is not important (here the difference between
attracting set and attractor matters: the latter being very often
fractal, cf.\ footnote 2 above). If the CH is valid for the GNS
systems, the FT of \cite{GalCoh95} follows {\it provided the
attractor is dense in phase space} and 
{\it provided the evolution is time reversible}.
The FT imposes strong restrictions, which we
illustrate below, on the form of the fluctuations of the contraction
of the surface elements on the attracting set (cf.\ eq.(\ref{tr})).

Let $t \mapsto S_t \uu$ be the time evolution for the initial velocity
field $\uu$ generated by the equations of motion on phase space, and
call $T$ the {\it total} simulation time, which must be adequately
longer than the characteristic time $\theta$ of the fluctuations of
$\sigma^{GNS}$. Assume that relaxation to a statistically stationary
state takes place in a time $t_0 \ll T$ with $t_0\gg\theta$ where
$\theta$ is a characteristic time for the evolution, {\it e.g.}
$\theta=1/\lambda_{max}$, and introduce the infinite time average of
the phase space contraction rate $\sigma^{GNS}$:
\be \langle \sigma^{GNS} \rangle = \lim_{T\to\infty} \frac{1}{T}
\int_0^T \sigma^{GNS}(S_t \uu) d t \,=\, \int \mu(d{\bf u}) 
\sigma^{GNS}(S_t \uu) .  \ee
If $T$ is large, we can subdivide the time interval $[t_0,T]$, into a
number of subintervals of length $(\zt+t_d)$, with $\zt\gg \theta$ and
$t_d$ a decorrelation time, separating the consecutive evolution
segments of length $\zt$. Consider the dimensionless quantities
\be
    \overline{\sigma^{GNS,\;\tau}}(i) =
    \frac{1}{ \tau\,\langle\sigma^{GNS}\rangle} 
    \int_{t_0+(i-1)(\zt+t_d)+t_d}^{t_0+i(\zt+t_d)}
    \sigma^{GNS}(S_t \uu)~d t ~,
\label{oversig}
\ee
with $i = 1, ..., (T-t_0)/(\tau+t_d)$ and arrange them in a histogram
which allows us to approximate the probability distribution $\pi^\zt$
of the values of $p=\overline{\sigma^{GNS,\;\tau}}$.  The total time
$T$ is empirically chosen large enough so that larger simulation
times and different initial fields produce differences in the
probability distributions that lie within the error bounds over a
range of $p$'s which is sufficient to perform meaningful tests.

Within the range of values of $p$ in which the number $n$ of events
per histogram bin is statistically meaningful,\footnote{Here, we
chose the range with $n \ge 10$ ($n$ being the number of events in a
bin), and we weighted the contribution of each
bin to the fit as inversely proportional to the associated error
bar. In this way, the cases with low statistical relevance (e.g. with $n
\approx 10$) have little influence on the computed $c(\zt)$'s and,
particularly, on its values extrapolated to large $\zt$.}  and for
sufficiently large $\zt$, it was observed (cf.\ \cite{RonSeg99}) that
the quantity
\be
 F(p;\zt) =  \frac{1}{\tau\,\langle\sigma^{GNS}  \rangle} \left[
  \log \left( \pi^\tau(p) \right) - 
  \log \left( \pi^\tau(-p) \right)
  \right] 
\label{finite0}
\ee
could be fitted by a function linear in $p$ with slope $c(\tau)$,
obtaining:
\be
\frac{1}{\tau\,\langle\sigma^{GNS} \rangle} \left[ \log \left(
\pi^\tau(p) \right) - \log \left( \pi^\tau(-p) \right) \right] =
c(\tau) p ~,
\label{finite}
\ee 
within the error bars.
If the conditions of the CH and of time reversibility are verified, one
can check whether or not $c(\tau)$ converges to a limit $c_\infty$, as
$\zt$ is increased. {\it If the attractor was really dense in phase
space,\footnote{Hence the whole phase space would be an attracting
set.}  then the validity of the FT would imply that $c_\infty$ exists
and, precisely, that $c_\infty=1$.} In such limiting case we refer to
Eq.(\ref{finite}) by calling it the ``fluctuation relation'' (FR).

However, the attractors of systems with strong dissipation often {\em
are not dense} in phase space. This is the case of the NS equations,
in which the attracting sets have a large open complement. In such
cases further assumptions are needed before the CH can be used to
derive interesting predictions. We shall consider the possibility that
there is proportionality between phase space contraction of volume
elements {\it around} the attracting set and contraction of surface
elements {\it on} the attracting set itself (which has dimension lower
than that of phase space).  This somewhat surprising possibility can
be related to (and predicted from) other assumptions which involve
geometric and dynamic hypotheses introduced in general in \cite{BG}
and applied to the GNS equations in \cite{GAFLU}.

\begin{figure} 
\centering
\psfig{figure=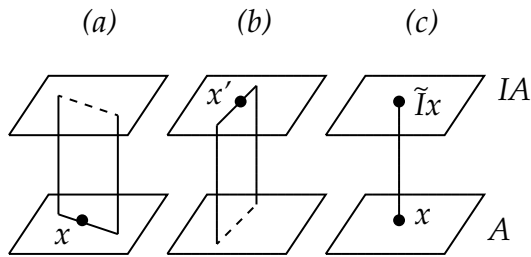,width=7.9cm,angle=0}
\caption{\small Illustration of axiom-C systems. The bottom squares
represent the attracting set $A$, while the top squares represent
the repelling set $IA$. The leftmost figure with vertical segments 
represents a point $x \in A$ with a piece of its stable manifold 
connetcing $A$ to $IA$. Similarly, the central figure represents a 
point $x' \in IA$ with a piece of its unstable manifold. The rightmost
figure represents the corresponding intersection of the stable manifold 
of $x$ with the unstable manifold of $Ix$, which associates a point
$\tilde{I}x$ in $IA$ with a point $x$ in $A$.}
\label{figaxC}
\end{figure}

To understand the hypotheses, first note that smoothness of the
attracting set and reversibility of the dynamics do not suffice to
deduce Eq.(\ref{finite}). Indeed, the image under time reversal of the
attracting set $A$ will be a repelling set $IA$ which, if the
attractor is not dense in phase space, is disjoint from it, cf.\
Fig.\ref{figaxC}.

Nevertheless the following may happen and be stable under perturbations of
the equations of motion.  Consider the stable manifold of the points
on the attracting set: part of this manifold will be on the attracting
set (which we suppose to be a smooth surface) and another part will stick
out, cf. (a) in Fig.\ref{figaxC}. Likewise the unstable
manifold of the points on the repelling set will be partly on it and
partly stick out, cf. (b) in Fig.\ref{figaxC}.

It is possible that the parts of the manifolds that stick out cross
each other transversally along a manifold that connects the attracting
and repelling sets, determining on each of them a single intersection
point, cf. the line in (c) in Fig.\ref{figaxC}. In this
way, a correspondence $\widetilde I$ between attracting set and
repelling set is constructed \cite{BG}.  By
construction the map $\widetilde I$ {\it commutes} with the time
evolution maps $S_t$.
The composition $I^*$ of $\widetilde I$ with the time reversal map $I$
of Eq.(\ref{tr}) leaves the attractor {\it invariant} and {\it
anticommutes} with time evolution restricted to it, {\it i.e.}
$$
S_t \cdot I^* = I^* \cdot S_{-t} \quad \mbox{on the attracting set.}
$$
A system verifying the above geometric properties, introduced and
discussed at some length in\cite{BG} and in \cite{Gal98,Galibro-i}, is
said to verify {\it Axiom C}. For such systems, time reversal symmetry
can be ``spontaneously broken'' as a control parameter ($R$ in
our case) varies, meaning that the attracting set $A$ becomes smaller
than the whole phase space, and distinct from the repelling set
$IA$. However, upon breaking, it spawns a new (lower) symmetry $I^*$
which acts on the attracting set and anticommutes with time
evolution. In a sense, in such systems time reversal {\it cannot be
broken} as it is always dynamically regenerated.

In Refs.\cite{BG,Gal98} this scenario was presented as the simplest
possible way in which time reversal symmetry could be spontaneously
broken via the splitting of the phase space of a reversible system
into an attracting set, a repelling set and a large region in between,
as in Fig.\ \ref{figaxC}. {\it Under this assumption}, the
CH holds for the dynamical system obtained by restricting the dynamics
to the attracting set. To such system the FT applies, and the
fluctuations of the contraction rate {\it of the area element} of the
attracting set obey the FR with an asymptotic slope $c_\infty=1$.

A further difficulty is that, usually, it is not possible to measure
the contraction rate of the surface of the attracting set because this
set is practically inaccessible dynamical object. However such a
measurement is not necessary if one can relate the area contraction on
the surface of the attracting set to the {\it total} phase space
contraction rate which is easily accessible, at least in
numerical experiments.

In Figs.\ref{n002n001},\ref{sovra18} the $2K-2$ nontrivial Lyapunov
exponents are arranged in pairs, each consisting of the
$k$--th largest exponent and the $k$--th smallest exponent (see 
\cite{Gal97} for a heuristic theoretical argument motivating 
this arrangement of our data). We imagine that 
the pairs with one positive and one negative exponent
pertain to the restriction of the evolution to the attracting
set.\footnote{ By the axiom C, i.e. by the consequent existence of a
time reversal symmetry on the attracting set, we know that on such set
half of the exponents is positive and half is negative, once the
vanishing exponent associated with the flow direction has been
discarded together with the other vanishing exponent associated with
the conserved quantity (energy, or enstrophy or any of the global
quadratic quantities $Q_m$, depending on the model). Arranging the
remaining $2K-2$ exponents in decreasing order: $\zl_1 \ge \zl_2 \ge
... \ge \zl_{2K-2}$, one can form the pairs $(\zl_k,\zl_k')$ of Figs.\
\ref{n002n001},\ref{sovra18}.} The other pairs, consisting of two negative
exponents, would then pertain to the directions of the stable manifold
that stick out of the attracting set.

Adopting the above view, we see that in a time $T$ and on the
attracting set the phase space contracts by $\sum^*
(\Lambda_{k,T}+\Lambda_{2K-1-k,T})$ where $\Lambda_{k,T}$ denotes the
``local contraction'' exponent of the surface of the attracting set
over a time interval $T$ around the phase point occupied by the system
at the beginning of the interval. Here, $\sum^*$ denotes the summation 
over the values of $k$ corresponding to pairs of exponents with one 
positive and one negative exponent. 
The total contraction of phase space is $\sum_1^K
(\Lambda_{k,T}+\Lambda_{2K-1-k,T})$. The two contractions are
asymptotic to $T \sum^*(\lambda_k+\lambda_{2K-1-k})$ and to $T
\sum^K_{k=1}(\lambda_k+\lambda_{2K-1-k})$, respectively, if
$\lambda_k$ denotes the $k$-th nontrivial Lyapunov exponent. Hence,
their ratio tends to 
\bea
c &=& \frac{\sum^*(\lambda_k+\lambda_{2K-1-k})}{\sum^K_{k=1}(\lambda_k+
\lambda_{2K-1-k})} = \nonumber\\
 &=& \lim_{T\to\infty}
\frac{\sum^*(\Lambda_{k,T}+\Lambda_{2K-1-k,T})}
{\sum^K_{k=1}(\Lambda_k+\Lambda_{2K-1-k})}=\lim_{T\to\infty} C(T) ~.
\label{pendenza}
\eea
If the ratio $C(T)$ is constant over the typical time scales over
which we observe the fluctuations of the phase space contraction,
there is proportionality between the phase space contraction and the
area contraction on the surface, and the proportionality constant is
precisely $c$ (which implies a weaker property than the one envisaged
in\cite{Gal97}: the latter in fact seems too strong and not verified
in the present case).  We have tested whether this is the case,
by taking as time scale $1/\zl_{max}$, the inverse of the largest Lyapunov
exponent.
The result is that, after a short transient, the quantity $C(T)$ 
is indeed practically constant 
(with fluctuation amplitudes of a few 
percent of its average value)
while the phase space contraction fluctuates
with {\it much larger} amplitudes. This is also reflected, for instance, in the
variances of the histograms of $C(T)$ with short $T$'s for the cases we checked,
which are more than one order of magnitutde smaller than the variances of the
corresponding histograms of the phase space contraction.
Therefore the FR with slope $1$ for the (not directly
accessible) surface contraction on the attracting set will imply a FR for
the {\it total} (measurable) phase space contraction with a slope $c$
given by Eq.(\ref{pendenza}).

\begin{figure}
\centerline{\hbox{
\epsfig{figure=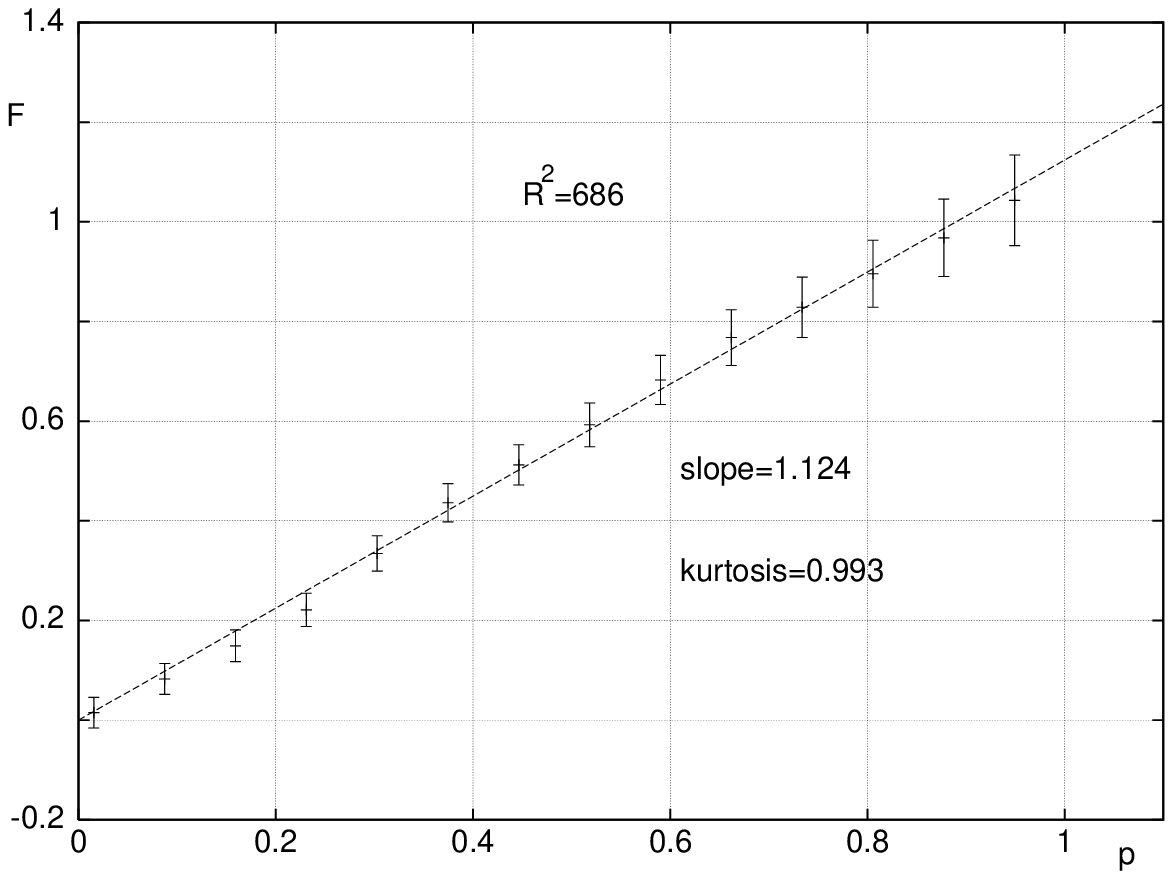,width=7.5cm}
\epsfig{figure=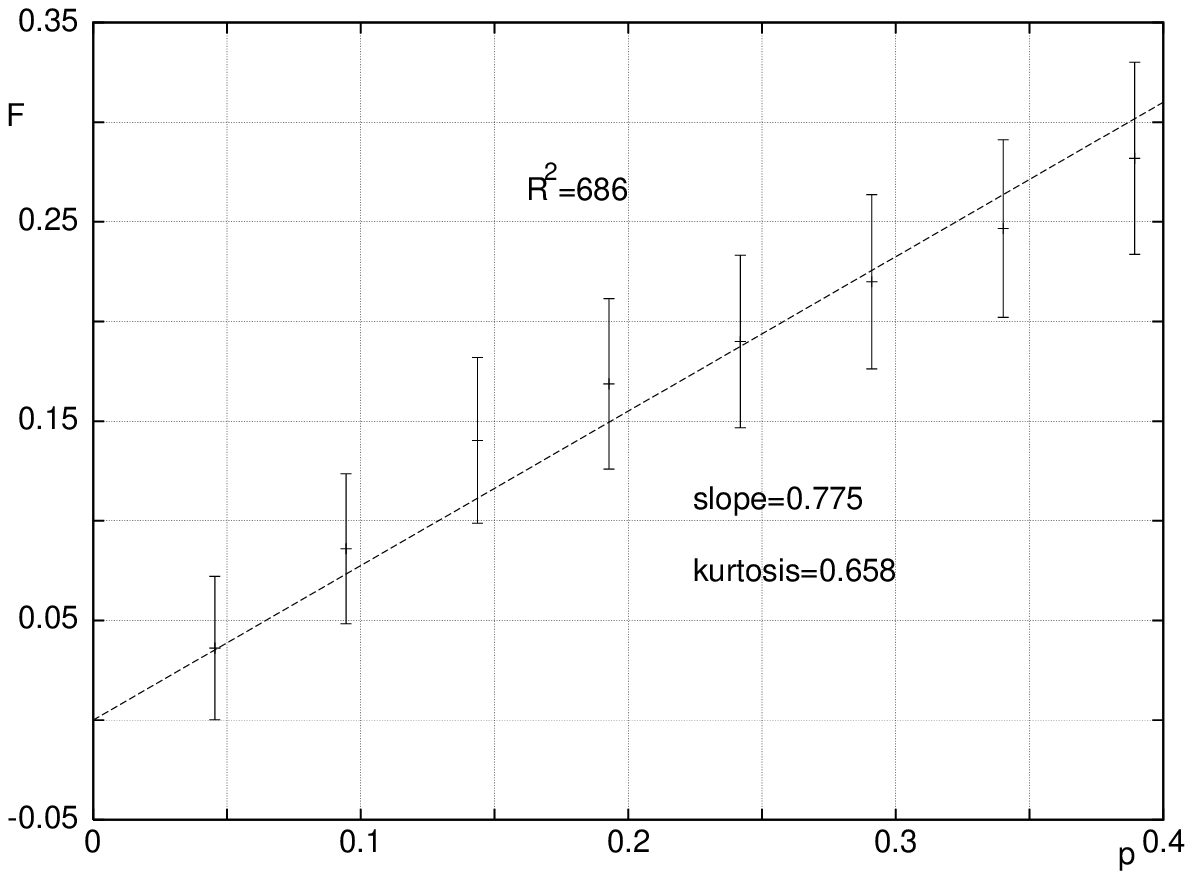,width=7.5cm}
}}
\caption{\small Computed values of $F(p;\zt)$, Eq.(\ref{finite0}),
for a GNS system
with fixed energy ($\ell=1$, $m=0$) and $R^2=686$.  The straight lines
interpolate the data, and the error bars represent the
statistical errors. 
The left panel has $\zt \approx 1.2$, while the 
right panel
has $\zt \approx 2.5$, in $1/\zl_{max}$ units.
In order to obtain the fit in Fig.\ref{cGNS} we repeated similar
measurements for various values of $\tau$.
Data refer to a
$24$--real modes truncation ($2K=24$). 
The data for $p<0$ can be inferred from the
symmetry $p \leftrightarrow -p$. The kurtosis of $\pi^\zt$ is defined by 
$\langle (p-1)^4 \rangle / \langle (p-1)^2 \rangle^2$, and equals $3$ if
$\pi^\zt$ is Gaussian.}
\label{verifica}
\end{figure}

\begin{figure}\centering
{\epsfxsize=10cm \epsfbox{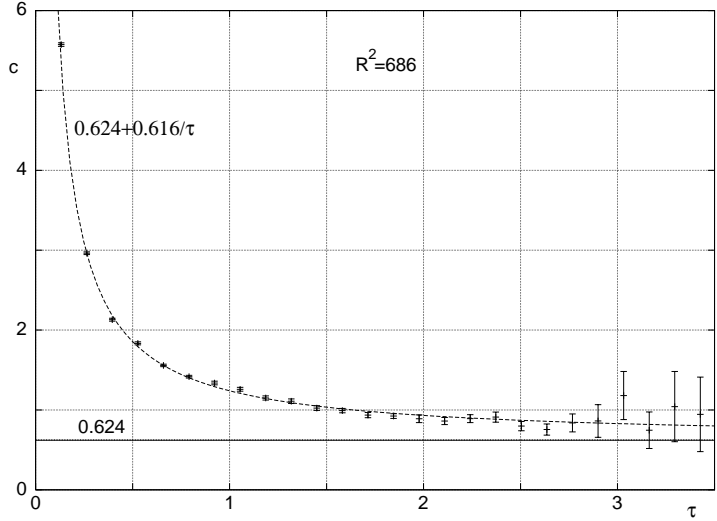}}
\caption{\small The value $c(\tau)$ for the GNS run with fixed energy
($\ell=1$, $m=0$), $R^2=686$. The times $\zt$, in the horizontal axis
are given in $1/\zl_{max}$ units. Data refer to a $24$--real modes
truncation ($2K=24$) and to values of $\tau$ to which already
correpsonds a linear graph of $F(p;\tau)$. The datum with smallest
$\tau$ has not been used in constructing of the fit (but it matches
it, nevertheless).}
\label{cGNS}
\end{figure}

If in Eq.(\ref{finite0}) we use the histograms for $\pi^\zt$ obtained
as outlined above, the result is well fitted by the
linear law (\ref{finite}). This is the case starting from $\zt$'s
comparable to $1/\zl_{max}$, and remains valid throughout the range of $\zt$'s
that we can consider. In this
range, the values of $c(\zt)$ can be computed through a least square
fit and, as suggested by the CH, the results can be
interpolated by a function like
\be c_\infty + A/\zt ~,
\label{extra}
\ee 
where $c_\infty$ is the asymptotic slope and $A$ is a constant: in
fact, in Anosov systems the limit of $\tau^{-1}\log \pi^\tau(p)$  is
reached with errors of order $\tau^{-1}$,  as $\tau\to\infty$
\cite{GalCoh95-0,GalCoh95}. In all cases we considered, such an
interpolation fits well our data. Figures \ref{verifica} -- \ref{cGNS}
illustrate that the above statements are valid even in the cases with 
small $R^2$. The reason for
the smallness of the $\zt$'s that we can consider can be understood by
looking at Fig.\ref{verifica} where, by comparing the graph on the
left panel (smaller $\tau$) with the one on the right panel (larger
$\tau$), we see that the range of $p$'s over which Eq.(\ref{finite0})
can be assessed rapidly decreases with $\zt$. This is due to the
decrease of the frequency of the negative fluctuations of
$\overline{\sigma^{GNS,\;\tau}}$, which rapidly becomes practically
zero as $\zt$ grows.

Nevertheless, the interpolation by $c_\infty + A/\zt$, where the
$c_\infty$ and $A$ are obtained from a least square fit, looks
convincing within our range of $\zt$'s and we therefore investigate
the consequences of assuming that it remains valid beyond this range.
In this case, one can compare the values of $c_\infty$ with those of
$c$ obtained from Eq.(\ref{pendenza}). The results are reported in
tables \ref{tabGNS} and \ref{tabGED}.  One realizes that, within the
accuracy of our numerical results, discussed more in detail in the
next section, there is agreement between the predictions of
Ref.\cite{Gal97} for $c$ and our values for $c_\infty$. Of course,
this is only a consistency check of the axiom C hypothesis which, as
mentioned above, can only be indirectly tested in our cases.

We also remark that the axiom C hypothesis leads to a simple
expression for the dimension of the attracting set, according to which
the ratio between the dimension of the attracting surface (on which
the attractor is dense) and the dimension of the phase space 
($2K-1$ in our cases because of the existence of the conserved
quantity $Q_0$) is the ratio $D_P=(2K-1-2n_-)/(2K-1)$, where $n_-$ is the
number of pairs with two negative Lyapunov exponents.
Obviously, the fractal
dimension of the attractor cannot be larger than the dimension
of the attracting surface. 
Therefore, under our assumptions on the nature of the attracting set,
if the Kaplan--Yorke dimension is a measure of the fractal dimension
of the attractor, its value divided by 
$2K-1$ should be $\le D_P$. It is therefore of some interest to 
check whether this holds in our case or not.

In most cases cases ({\it e.g.} in the cases of Fig.\ref{n002n001}) there
are three exponents very close to $0$ and they are affected by a large 
relative error (see comments in sect. 5 below). We interpret as $0$ the two 
of smallest absolute value (as we know that two exponents necessarily vanish).
Although the signs of the second and third smallest exponents are stable with  
the length of our simulations, some uncertainty remains, at times, 
on which of the two approximates a zero exponent, and which approximates a
finite exponent. For this reason, the values $n_-$ reported in brackets 
in Tables \ref{tabGNS} and \ref{tabGED} are uncertain by $1$, and this
implies an uncertainty also on the calcualtions of $c$ and
$D_P$ (cf.\ point 3 of Section 6 below).

Table \ref{tabGNS} shows that in our system the axiom C prediction and
the KY value do not seem really compatible.
The comparison of the dimensions and its consequences has
been suggested by F. Bonetto. The incompatibility between the
axiom C and the KY methods for determining the fractal dimension of an
attractor seems to be quite general and to apply to rather
different kinds of chaotic motions.
\begin{table}
\begin{center}
\begin{tabular}{||c|c|c|c|c||} \hline \hline
$ R^2$ &$ c $    &   $c_\infty$     & KY     & $D_P$        \\
\hline					    
  556  &0.534    & 0.535 $\pm$ 0.008 & 0.734  & 0.565 ($n_-=5$)\\
\hline					       
  638  &0.436    & 0.421 $\pm$ 0.020 & 0.610  & 0.478 ($n_-=6$)\\
\hline					       
  686  &0.630    & 0.624 $\pm$ 0.013 & 0.848  & 0.652 ($n_-=4$)\\
\hline					       
  800  &0.639    & 0.614 $\pm$ 0.025 & 0.830  & 0.652 ($n_-=4$)\\
\hline					       	   
 1250  &0.639    & 0.621 $\pm$ 0.024 & 0.845  & 0.652 ($n_-=4$)\\
\hline					       
 2222  &0.834    & 0.717 $\pm$ 0.011 & 0.914  & 0.826 ($n_-=2$)\\
\hline					       
 5000  &0.842    & 0.839 $\pm$ 0.015 & 0.939  & 0.826 ($n_-=2$)\\
\hline \hline
\end{tabular}
\caption{\small Properties of the GNS dynamics with fixed energy
($\ell=1$, $m=0$). The value of $c$ is obtained from
Eq.(\ref{pendenza}) using the measured Lyapunov exponents. The value
$c_\infty$ is obtained by extrapolation of the data as in Fig.\
\ref{cGNS}. The column labeled by
KY, gives the Kaplan-Yorke dimension computed from our measured
Lyapunov exponents.
The column labeled by $D_P$ is the ratio $(2K-1-2n_-)/(2K-1)$ where
$n_-$ is the number of pairs with two negative Lyapunov exponents.
Data refer to a $24$--real modes truncation ($2K-1=23$). 
Our estimate of $n_-$, given in brackets in the column of $D_P$,
affects the calculations of both $c$ and $D_P$.}
\label{tabGNS}
\end{center}
\end{table}

The same questions addressed here have been investigated in the case
of the GED equations, obtaining the results reported in Table
\ref{tabGED}.
\begin{table}
\begin{center}
\begin{tabular}{
||  c  |c   | c       |            c      | c ||} \hline \hline
$ R^2$ & $c$ & $      c_\infty$    &  KY    & $D_P$        \\
\hline 					   
  408  & 0.911 & 0.823  $\pm$ 0.009 & 0.943  & 0.913 ($n_-=1$)\\
\hline 					   
  556  & 0.813 & 0.812  $\pm$ 0.006 & 0.957  & 0.826 ($n_-=2$)\\
\hline 					   
  800  & 0.910 & 0.910  $\pm$ 0.007 & 0.983  & 0.913 ($n_-=1$)\\
\hline \hline
\end{tabular}
\caption{\small Properties of the GED dynamics with fixed energy
($\ell=0$, $m=0$). The meaning of the columns is the same as in
table \ref{tabGNS} above. 
Data refer to a $24$--real modes truncation ($2K-1=23$).}
\label{tabGED}
\end{center}
\end{table}

It should be noted that the {\it a priori} pairing assumption
which, together with axiom C, allowed us to interpret and organize our
data, does not have firm theoretical grounds other than the discussion
in \cite{GAFLU}. Therefore the above analysis is highly conjectural
and finds its main justification in its capability to fit the data.

\section{Technical coments on the simulations. Accuracy.}
The calculations summarized by Figs.\ref{n002n001},
\ref{verifica}-\ref{cGNS}, and by Tables \ref{newtab}--\ref{tabGED},
all refer to cases with rather small $R$ and $N$ compared to the cases
studied in Ref.\cite{RonSeg99}. Small $R$ is needed to obtain values
of $c_\infty$ sufficiently small to be distinguished clearly from $1$,
while small $N$ is needed to compute with reasonable accuracy the
Lyapunov spectra.  The actual values $R$ and $N$ that we chose
constitute a compromise, but the result is that the frequency of the
negative fluctuations, needed for a test of the fluctuation formula,
is quite low, and very rapidly it becomes negligible for growing
$\zt$'s, even in long simulations. Given our present computer
facilities, our standard simulation was of order of $10^8$ time steps,
with sizes $DT$ of order between $10^{-6}$ and $10^{-4}$,
corresponding to times $T$ of order between $10^3$ and
$10^5$ in $1/\zl_{max}$ time units.

Such $T$'s suffice for calculations of the average of quantities like
the $Q_m$'s, and to build with some accuracy the histograms of the
fluctuations of $\overline{\zs^{GNS,\tau}}$ for relatively small
$\zt$'s. The accuracy is reflected by the error bars placed on the
computed quantities, like in Figs.\ref{verifica}-\ref{cGNS}, which
cover the range of $\zt$'s beyond which the negative fluctuations are
too rare to have any statistical significance (cf.\ footnote 8). These
bars represent the statistical errors, obtained from the error bar on
the histogram for $\pi^\zt$, i.e.\ $\pm \sqrt{n}$ on a bin, if $n$ is
the number of counts in that bin.

The times $T$ are also sufficient for an accurate evaluation of the
largest Lyapunov exponents. Indeed, out of $24$, the largest (in
absolute value) $20$ exponents have an uncertainty ranging from $25\%$
down to $0.06\%$ of the computed value, for the largest one.  The
uncertainty on the remaining $4$ (smallest) exponents at times exceeds
$100\%$ of the computed value. The uncertainty on one exponent has been
evaluated as the amplitude of the largest fluctuation of the computed
value of that exponent in the second half of a simulation of length
$T$. In all cases, the error bars do not exceed 
the symbols size, if drawn in figures such as Fig.\ref{n002n001}.  
Nevertheless, at times, it  is not obvious 
which two, of the three smallest exponents, vanish and which does not.
This makes $n_-$ uncertain by one unit. In particular, in the GNS case with 
$R^2=2222$ and in the GED case with $R^2=408$, $n_-$ could have been taken
to be $3$ and $2$, respectively, rather than $2$ and $1$ as reported 
in Tables \ref{tabGNS} and \ref{tabGED}. In fact, the third and the second
smallest exponents are both quite small and close in absolute value, but the 
third smallest is positive while the second smallest is negative. Therefore,
considering zero the third smallest exponent, rather than the second, makes 
$n_-$ grow by $1$.

The slopes $c(\zt)$, from which the values $c_\infty$ of Tables
\ref{tabGNS} and \ref{tabGED} have been extrapolated, were computed by
fitting our data to Eq.(\ref{finite}) and by using our histograms of
$\overline{\zs^{GNS,\tau}}$ for $\pi^\zt$.
The limiting slopes $c_\infty$ and its error bar reported in Tables 
 \ref{tabGNS} and \ref{tabGED}
have been computed by a least square fit the function 
$c_\infty + A/\zt$ to the $c(\zt)$'s.

\section{Concluding remarks}
\noindent
{\bf 1.} The results on the equivalence between NS and GNS dynamics
obtained in Ref.\cite{RonSeg99} have been strengthened through
the analysis of the Lyapunov spectra of NS and GNS systems, albeit at 
very low
spectral resolution. Similar results were obtained for spectral
resolutions with up to $190$ modes.

\vskip 5pt \noindent
{\bf 2.} The results of Ref.\cite{RonSeg99} on the validity 
of the FR for the GNS dynamics have been strengthened. 
Unfortunately, the
relatively high dissipation of the systems considered here
did not allow us to study the fluctuations of 
$\overline{\zs^{GNS,\tau}}$, with 
$\zt >> 1/\zl_{max}^+$, as desired, due to lack of appreciably frequent
negative fluctuations 
at such $\zt$'s. Nevertheless, our data are well fitted by $c_\infty + A/\zt$,
in the whole ranges of $\zt$ we considered,
in accord with the validity of the CH.

\vskip 5pt \noindent {\bf 3.} The Lyapunov spectra have been used to
compute the values $c$ of Eq.(\ref{pendenza}) reported in Tables
\ref{tabGNS} and \ref{tabGED}. These values
equal the corresponding quantities $c_\infty$ in agreement with the  
axiom-C hypothesis, except for the GNS case with $R^2=2222$ and for the
GED case with $R^2=408$. The uncertainty on $c$ looks more
pronounced than the uncertainty on $c_\infty$, because of the
uncertainty on the number of positive Lyapunov exponents arising from
the fact that often one of them is too close to $0$.\footnote{Two
exponents have to vanish: the one corresponding to the conserved
quantity $Q_m$, and the one corresponding to the direction of the
flow. However there are other exponents very close to zero, and one
faces, therefore, the difficult task of distinguishing numerically the
small (positive or negative) exponents from the vanishing ones.} The
effect of this uncertainty on $c$ is of about $9\%$, since it affects
by $1$ the number of pairs appearing in the numerator of
Eq.(\ref{pendenza}), and the contributions of the different pairs do
not fluctuate much around their average value. In particular, in the 
GNS case with $R^2=2222$ and in the GED case with $R^2=408$, taking
$n_-=3$ and $n_-=2$, respectively, leads to a close agreement
between $c$ and $c_\infty$ also in these two cases.

\vskip 5pt \noindent {\bf 4.} Our codes, considered as dynamical
systems in their own, seem to verify the axiom C, within the time
scales of our simulations. Whether this
property can be inferred to hold for the dynamics of real fluids is
not clear, because of the low spectral resolution we have
used. 

\vskip 5pt \noindent {\bf 5.} Global fluctuations of macroscopic
systems can hardly be observed, although the experiments of
Refs.\cite{CL,GGK} attempt at studying precisely this kind of
fluctuations. However, as argued in the Appendix, the (observable) local
fluctuations seem to obey laws similar to those of the global fluctuations
discussed here.

\vskip 5pt \noindent {\bf 6.}  The extrapolation hypothesis
Eq.(\ref{extra}) leads to a slope $c_\infty$ which has been fitted to
a value $<1$. If confirmed by further experiments, this implies that the
stochastic fluctuation theorems which lead to $c_\infty=1$, such as
those recently proposed in \cite{CM99}, would require extra ideas to
be adapted to the description of the fluctuations of our deterministic
dynamical systems. 

\section*{Appendix: Remarks on local fluctuations and on spatio--temporal
chaos.}
Recently, local versions of the FT have been considered to overcome
the problem that the large global fluctuations which are the object of
the original FT cannot be observed in real macroscopic systems. One
possible approach has been outlined in Ref.\cite{GGlocal}, where a
fluctuation relation similar to (\ref{finite}) has been derived for
the fluctuations of local quantities, in an infinite chain of weakly
interacting chaotic maps, which is a paradigmatic system exhibiting
spatio--temporal chaos. In particular, let $V_0$ be a finite
region of the chain centered at the origin, $T_0 > 0$ be a time
interval, and define $\eta_+$ and $p$ as
\be \eta_+ = \lim_{V_0,T_0 \rightarrow
\infty} \frac{1}{|V_0| T_0} \sum_{j=0}^{T_0-1} \eta_{V_0} (S^j x) ~,
\quad p = \frac{1}{\eta_+ |V|} \sum_{j=-T_0/2}^{T_0/2} \eta_{V_0} (S^j
x) ~,
\label{etaplus} 
\ee
where $V = V_0 \times T_0$, and $\eta_{V_0}(x)$ is a properly chosen
quantity which in \cite{GGlocal} is called  the {\em entropy production rate in
$V_0$} and is identified with the phase space contraction rate in $V_0$.
The following relation was obtained for the steady state 
probability distribution of $p$:
\be
\pi_V(p) = e^{\zeta(p)|V| + O(|\partial V|)} ~, \quad
\mbox{with } \frac{\zeta(p) - \zeta(-p)}{p \eta_+} = 1 \quad \mbox{and }
|p| < p^* ~,
\label{pilocal}
\ee
where $|\partial V|$ is the size of the boundary of $V$, $p^* \ge 1$ and
$\zeta$ is analytic in $p$. This is perfectly analogous to the original
FT, except for the boundary term $|\partial V|$ whose influence should
diminish with growing $V$.

The question of the validity of relations analogous to
Eqs.(\ref{etaplus},\ref{pilocal}) in 2-dimen\-sional fluids is beyond
the scope of the present paper.  Nevertheless, it seems worthwhile
to report our encouraging preliminary results  on the validity of a
local fluctuation relation for the NS dynamics. In fact, these results
make plausible a linear relation with slopes smaller than $1$, and indicate that
in the NS case, in
which dissipation occurs uniformly throughout the system, the local
fluctuations may obey laws very similar to those obeyed by the global
fluctuations considered in the previous sections. The interest lies in
the possibility that, in agreement with the EC, the local fluctuations
can be observed also in the NS dynamics, despite their
irreversibility.

The first study of a local fluctuation relation to obtain the value of
$c_\infty$ is in \cite{BL01}, and the problem has raised many
discussions since.

Let us consider the NS equation on a periodic cell of side $2 \pi$,
and let $V_0 = [-L/2,L/2] \times [-L/2,L/2]$ be a box with side $L \le
2 \pi$. We define the power dissipated in $V_0$ by the forcing field
${\bf f}$ at time $t$ as: 

\be \hat{p}_L(t) = \int {\bf f}(x) \cdot \U(\xx,t) \chi_{_{V_0}}(\xx)
~d \xx ~,
\label{piellhat}
\ee
where $\chi_{_{V_0}}$ is the characteristic function of $V_0$ which,
in spectral form, is: $\chi_{_{V_0}}(\xx) = \sum_{\kk} \chi_{_\kk}(L)
~ e^{i \kk \cdot \xx}$.
Defining local vorticity, local palinstrophy and $\za_L$,
respectively, as
\bea 
Q_{1,L}(t) &=& \int \zw^2(\xx,t) \chi_{_{V_0}}(\xx) ~d \xx ~, 
\nonumber \\
Q_{2,L}(t) &=& \int (\nabla \zw)^2(\xx,t) ~\chi_{_{V_0}}(\xx) ~d \xx ~, 
\quad \za_L(t) = \frac{\hat{p}_L(t) }{Q_{1,L}(t)} ~,
\label{Qloc}
\eea 
with $\zw = \nabla \times \U$, we propose to take $\eta_{V_0}(\uu)$ as
proportional to the work per unit time performed by the external force
within the region $V_0$ divided by the corresponding local vorticity,
i.e. we propose to take $\eta_{V_0}(\uu)$ proportional to $\za_L$.
This is suggested by the facts that such a quantity has a physical
meaning, and that the phase space contraction rate can in many systems
be identified with the entropy production rate. In fact, $Q_{1,L}$ can
be called a ``local temperature'' being a quantity whose value
measures the variations of the velocity field around its
average. A possible alternative definition is a local version of
Eq.(\ref{pscr}), like (for instance)

\be \hat{\sigma}_{L} = 2 \left(\left({L\over 2\pi}\right)^2 \sum_{|k_i| \le N} 
|k_i|^2 ~ - ~ {Q_{2,L}\over Q_{1,L}} \right)\,
\cdot\,\alpha_{L} + {\int \nabla f\,\nabla \omega ~\chi_{_{V_0}}~d\xx
\over Q_{1,L}} ~.
\label{pscr-local}
\ee 
However, several other local versions of Eq.(\ref{pscr}) could be defined:
{\it e.g.} the last integral could be replaced by  
$\int \nabla^2 f\,\omega ~\chi_{_{V_0}} d\xx$ or by
$\int f\,\nabla^2 \omega ~\chi_{_{V_0}} d\xx$.

Since we approximate the various quantities introduced above by means
of truncated Fourier expansions, $L$ can only take the values $l
\pi/(N-1)$ with $l=1, ..., 2(N-1)$.

Studying the behaviour of $\zs_L = \hat{\zs}_L/ \langle \hat{\zs}_L
\rangle$, $\langle \hat{\zs}_L \rangle$ being the infinite time
average of $\hat{\zs}_L$, one realizes that $\hat{\zs}_L$ is
essentially proportional to $\za_L$, with proportionality constant
equal to the ratio $\langle \hat{\zs}_L \rangle/\langle {\alpha}_L
\rangle$, because the sum in (\ref{pscr-local}) dominates over the
other terms already at moderately large $L$.  In
particular, we performed a number of simulations with $N=7$ (i.e.\
$84$ complex Fourier modes) and $R^2 = 2048$, and we found that
$\hat{\za}_L/\langle\hat{\za}_L\rangle$ and
$\hat{\zs}_L/\langle\hat{\zs}_L\rangle$ are practically
indistinguishable from each other for $L \ge 5\pi/6$. The statistics
of the fluctuations of $\hat{\zs}_L$ was built from a run of $1.36
\cdot 10^8$ time steps, corresponding to $2.4 \cdot 10^4$
in $1/\zl_{max}$ time units.

\begin{figure}
{\epsfxsize=6.5cm \epsfbox{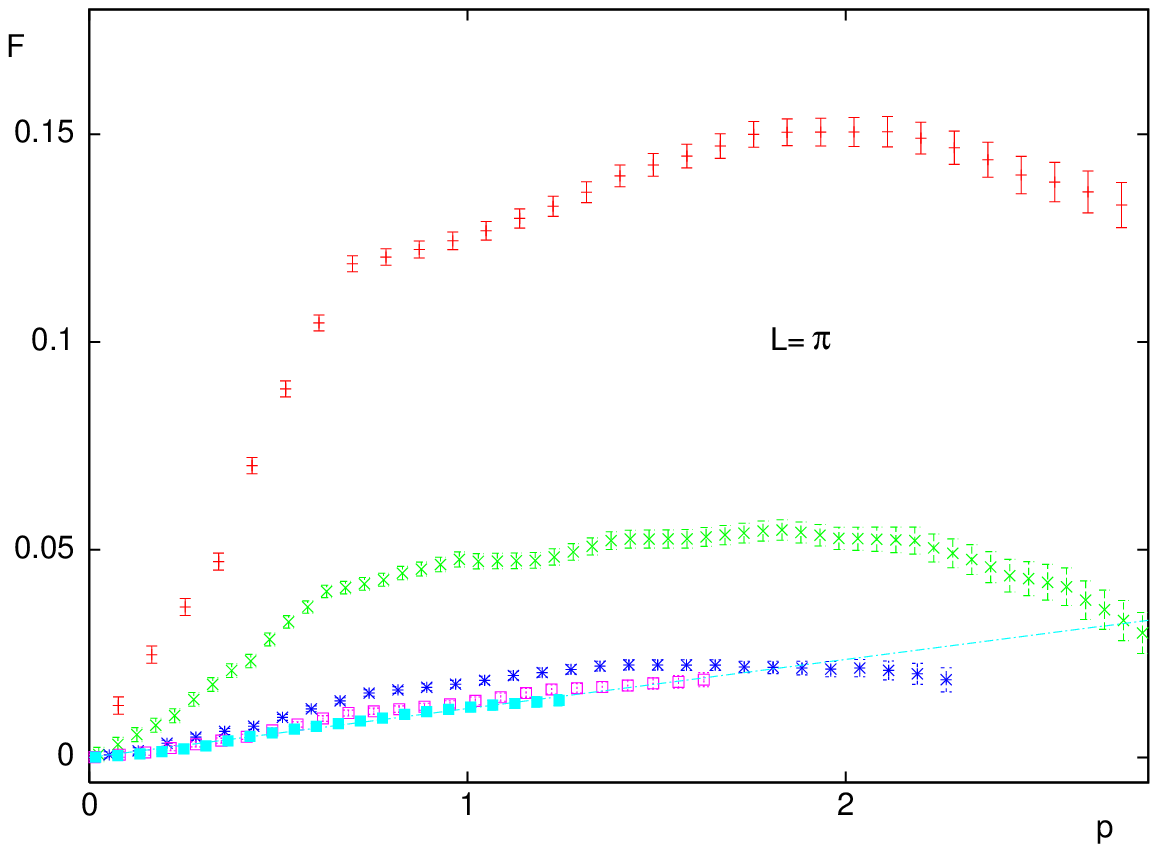}}
{\epsfxsize=6.5cm \epsfbox{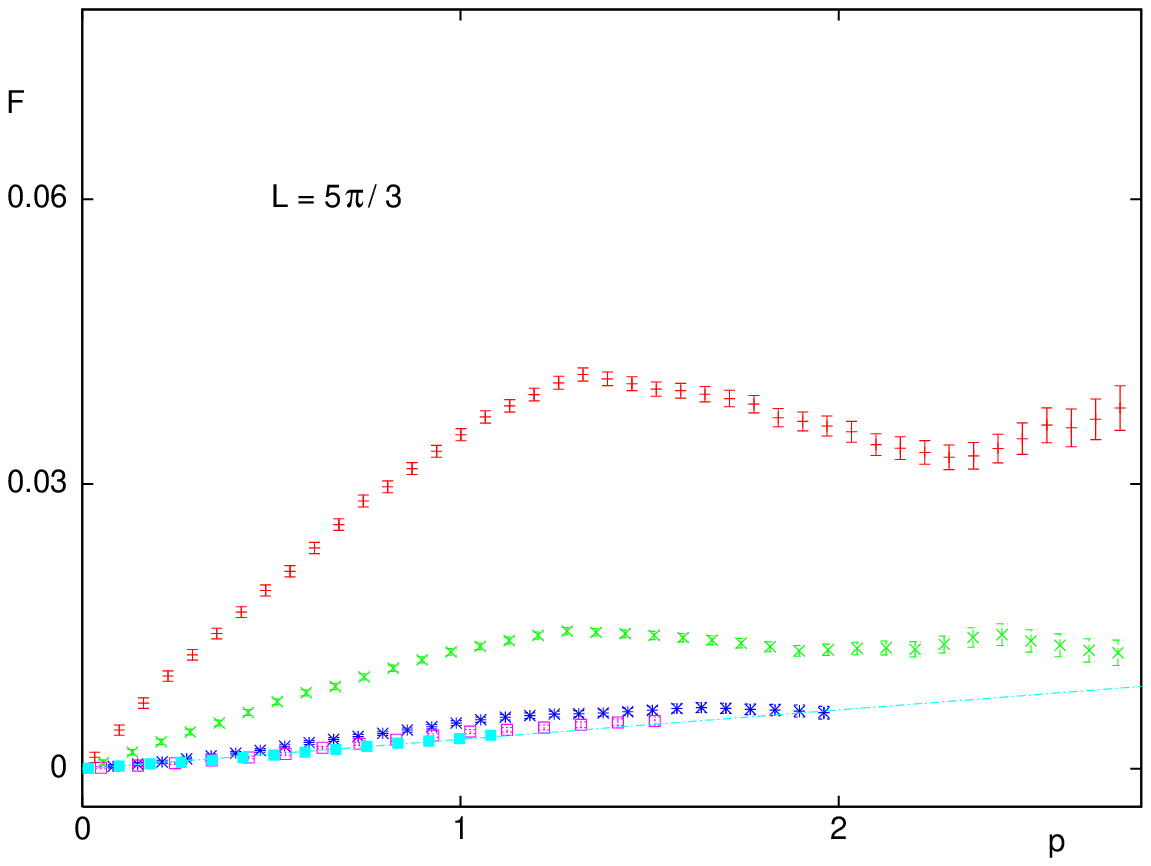}}
\caption{\small Numerically computed values of $F(p;V)$, and
corresponding error bars, for $L=\pi$ (left) and $L=5\pi/3$
(right). In each panel we have $5$ curves, corresponding to $T_0
\approx 0.58,1.75,5.27,7.81,10.74$ (in $1/\zl_{max}$ units), ordered
clockwise with growing $T_0$.  Only three curves are clearly
distinguishable because the three of highest $T_0$ practically overlap
in their range of coexistence, which shrinks with increasing $T_0$.}
\label{l4flut}
\end{figure}

\begin{figure}
{\epsfxsize=6.5cm \epsfbox{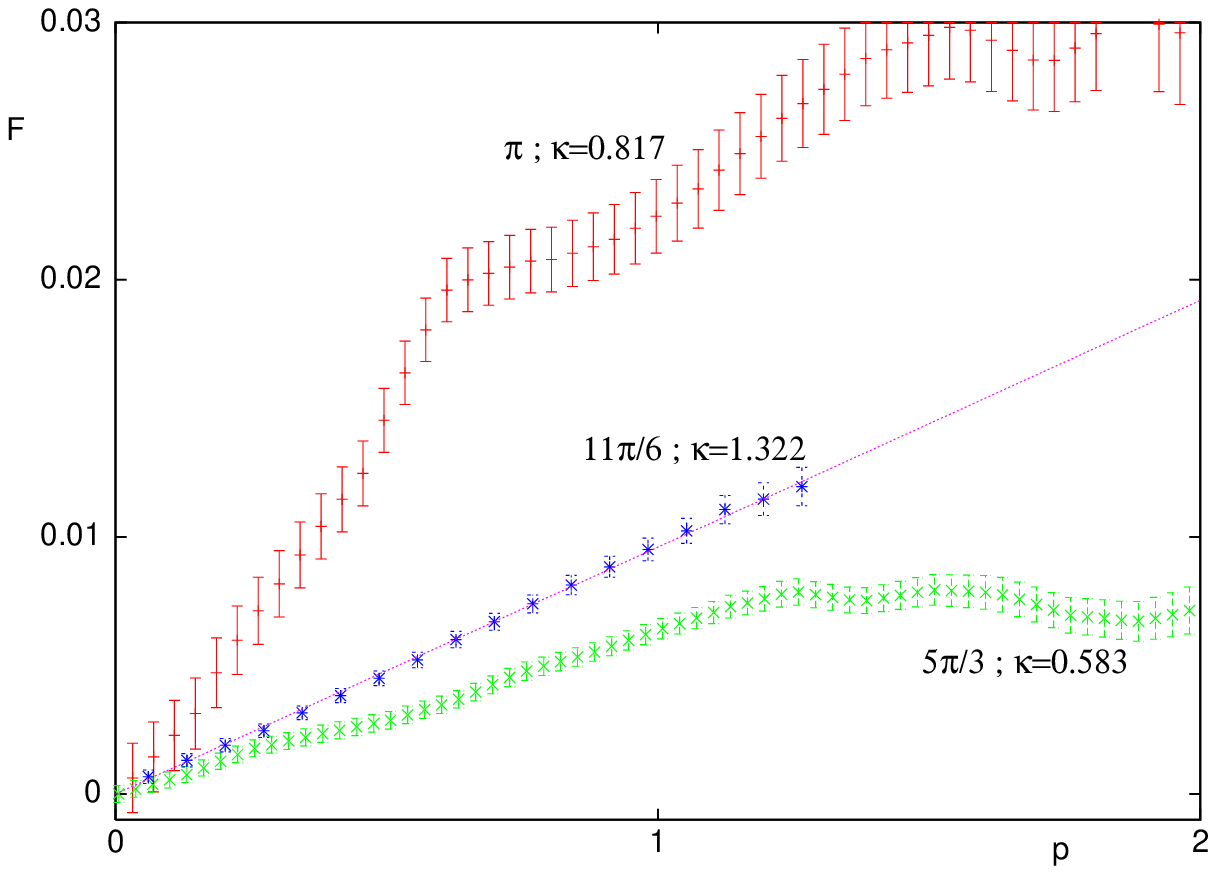}}
{\epsfxsize=6.5cm \epsfbox{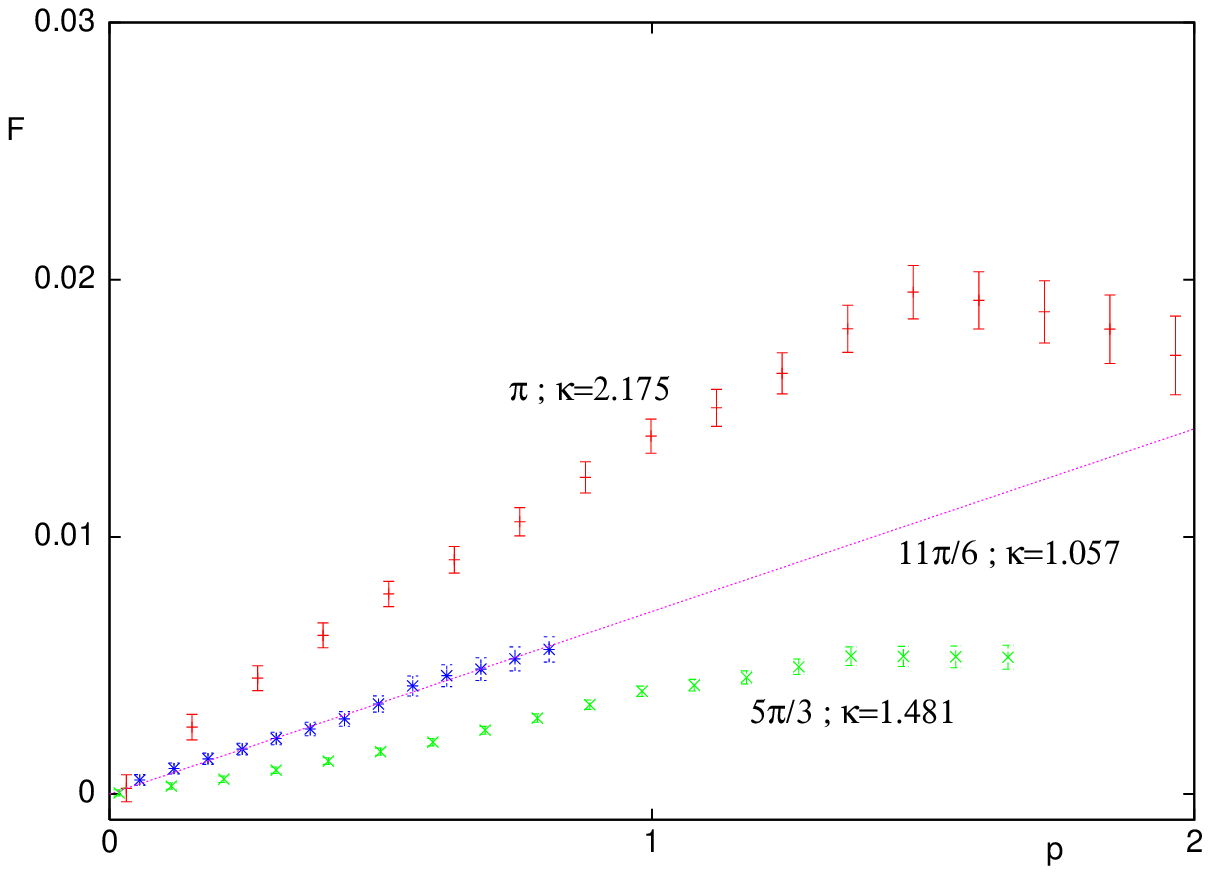}}
\caption{\small Values of $F(p;V)$, and
corresponding error bars, for $T_0 \approx 3.7$ (left) and
$T_0 \approx 7.0$ (right) in
$1/\zl_{max}$ units . In each panel we have $3$ curves,
corresponding to $L=\pi,5\pi/3,11\pi/6$, and the values of $\kappa$
represent the kurtosis of the corresponding PDFs.}
\label{f928}
\end{figure}

As in the case of global fluctuations (cf.\ Fig.\ref{verifica}, and
Ref.\cite{RonSeg99}), we found that the PDFs 
of the fluctuations of the average of $\zs_L$ over times $T_0$ are not
Gaussian (cf.\ Table \ref{kurto}, and Fig.\ref{f928}), and 
furthermore that the quantity 
\be F(p;V) = \frac{1}{T_0 \langle \hat{\zs}_L \rangle} \log
\frac{\pi_V(p)}{\pi_V(-p)} \quad \mbox{with} \quad p = \frac{1}{T_0
\langle \hat{\zs}_L \rangle} \int_t^{t+T_0} \hat{\zs}_L(s) d s
\label{FFlocal} 
\ee 
is not linear in $p$ at small $V$ (i.e.\ small $L$, or small
$T_0$), but it appears to be better and better approximated by a straight
line of slope $c(V)$ as $V$ grows.  Moreover, the slopes of the linear
fits of the data appear to converge to values $c_L$ for $T_0
\rightarrow \infty$.  Figure \ref{l4flut} illustrates the latter
convergence at fixed box size $L$ and growing $T_0$, showing that a linear
fit of the data becomes acceptable at sufficiently large $T_0$, and
suggesting that the slope of the fitting line may converge to a limit
value.
\begin{table}
\begin{center}
\begin{tabular}{
||  r  |  c  |  c ||} \hline \hline
$T_0$ & $L=\pi$ & $L=5\pi/3$\\
\hline 					   
 0.58  & 2.002 & 1.249 \\
\hline 					   
 1.75  & 1.065 & 1.476 \\
\hline 					   
 5.27  & 1.532 & 1.120 \\
\hline 					   
 7.81  & 1.215 & 1.448 \\
\hline 					   
10.74  & 0.928 & 1.066 \\
\hline \hline
\end{tabular}
\caption{\small Kurtosis of $\pi_V$ for the cases of Fig.\ref{l4flut}.
These values show that such PDFs are not Gaussian.}
\label{kurto}
\end{center}
\end{table}

\begin{figure}
{\epsfxsize=6.5cm \epsfbox{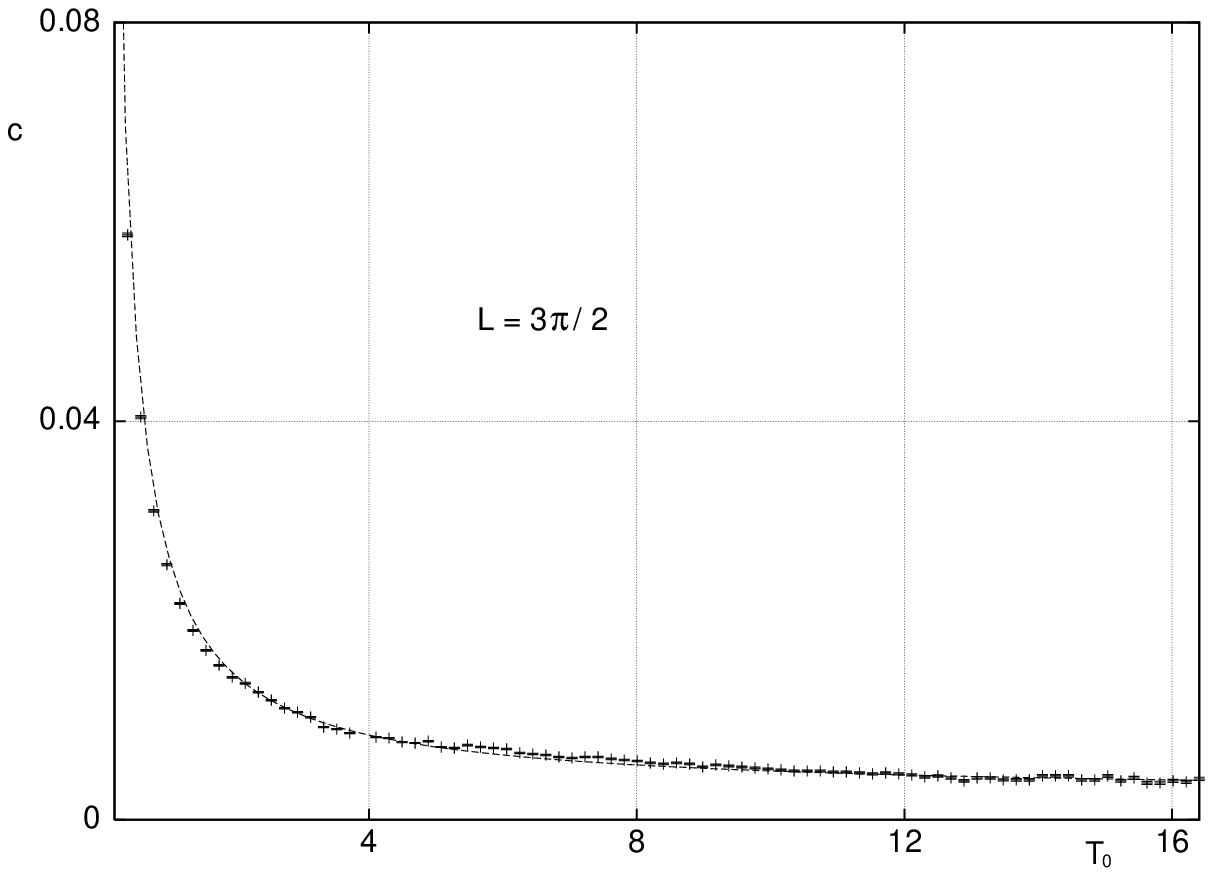}}
{\epsfxsize=6.5cm \epsfbox{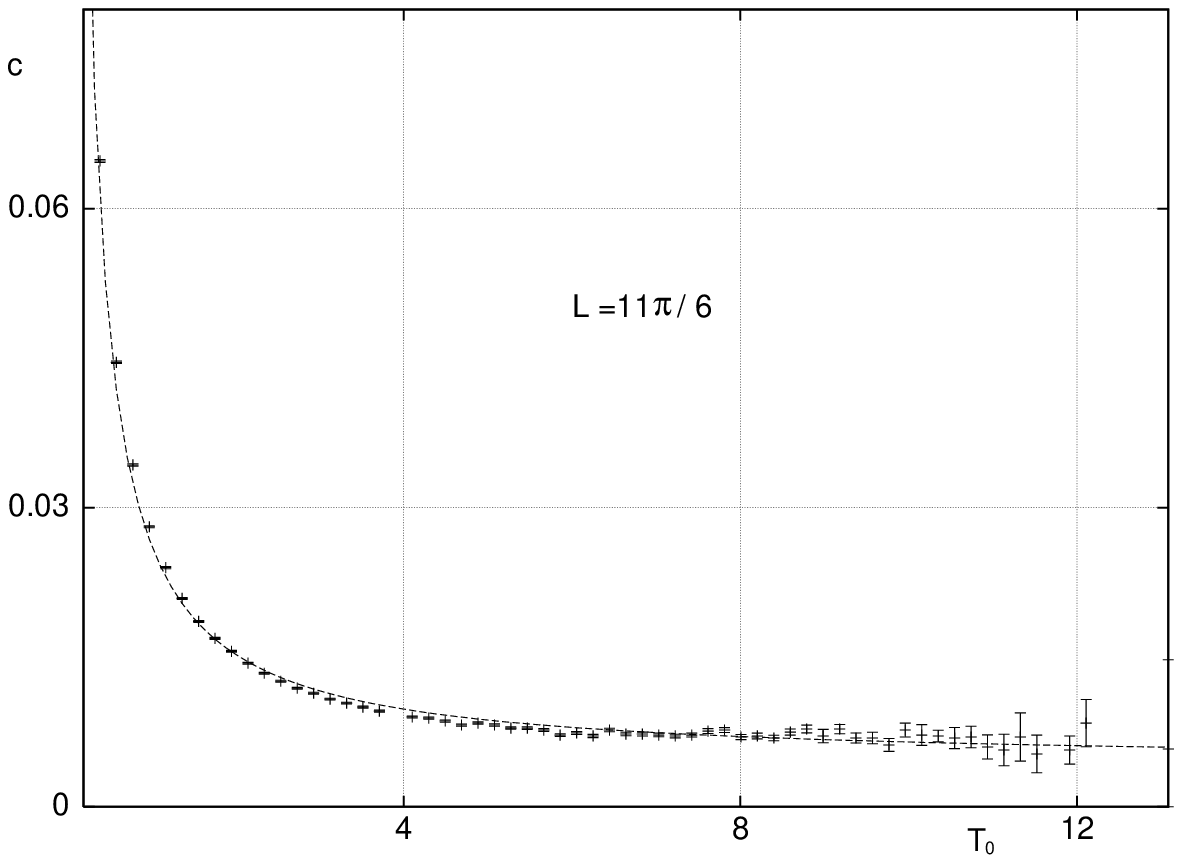}}
\caption{\small Slopes $c(V)$ for fixed $V_0$ as a function of $T_0$
(in $1/\zl_{max}$ time units).  
The left panel has $L=3\pi/2$ and the data are interpolated by $c =
0.0025+0.024/T_0$. The right panel has $11\pi/6$ and the data are
interpolated by $c = 0.0043+0.022/T_0$.}
\label{gcf48}
\end{figure}

Figure \ref{f928} shows that the same holds if $T_0$ is fixed and $L$
grows, indicating that the nonlinear tails of $F$, may possibly be
ascribed to non negligible boundary contributions, which become less
important at larger $L$.  In fact, we notice that, for any $V$,
$F(p;V)$ is linear in a given neighborhood $I_0$ of $p=0$, even if strong
nonlinearities characterize $F$ outside $I_0$. As $V$ grows, it seems
that the linear region persists, although it changes slope, while the
nonlinear tails of $F$ gradually disappear. Therefore, our data for $F$ justify
a test of the local fluctuation formula
(\ref{pilocal}), in analogy with the theory of Ref.\cite{GGlocal}. The
result is summarized in Fig.\ref{gcf48}, where the slopes $c$ of the
linear fits of our data, and their variation
with growing $\zt$ are portrayed.  There, we observe that our results for
the local fluctuations are consistent with the EC and EC$'$, and with
linear relations with limit slopes smaller than $1$. Slopes $<1$ have
also been previously found for local fluctuations in rather different
systems, \cite{BL01}.

We conclude by remarking that:

\noindent
{\bf a.} Our definition of the local phase space contraction rate
(\ref{pscr-local}) equals (up to a scaling factor) the physically measurable
quantity $\za_L$ for sufficiently large $L$, which in our case amounts to
$L=l\pi/6$ for $l=5,6,...,12$. 

\noindent
{\bf b.} The fluctuations of the local quantities $\hat{p}_L$ and
$\hat{p}_L/Q_{0,L}$, with $Q_{0,L}=\int \U^2(\xx,t) \chi_{_{V_0}}(\xx)
~d \xx$, behave in a perfectly analogous way as the fluctuations of
$\hat{\zs}_L$. The latter remark is important because in the
experimental work of Ref.\cite{GGK} the quantity which is measured to
attempt a check of the FR is essentially the work $\hat{p}_L$ done by
the external forces on a small region $V_0$ per unit time, rather than
the same quantity divided by an effective temperature.

\noindent
{\bf c.} The fluctuation laws tested in this Appendix concern the
({\it irreversible and constant friction}) NS equation, and not
the ({\it reversible and fluctuating friction}) GNS dynamics. This 
is consistent with the validity of the EC,EC$'$.
%

\section*{Acknowledgements}
We are grateful to S.\ Ciliberto and C.P.\ Dettmann
for highly stimulating discussions. LR gratefully acknowledges partial
support from GNFM. We are particularly indebted to
F.\ Bonetto for suggesting several improvements and for clarifying 
the meaning of the pairing rule, which resulted in the 
correction of an error in the manuscript. GG and LR thank B.\ Dorfman, 
P.\ Gaspard, R.\ Klages,  H.\ van Beijeren and the Max Planck Institute  
for Physics of Complex Systems, Dresden, for their generous and
warm hospitality
during the conference ``Microscopic Chaos and Transport in Many-Particle
Systems''.

\bibliography{biblio} \bibliographystyle{unsrt}
\end{document}